\documentclass[aps,pra,twocolumn,amsmath,amssymb,superscriptaddress]{revtex4-2}
\usepackage{times}
\usepackage[pdftex]{graphicx}
\usepackage{dcolumn}
\usepackage{bm}
\usepackage{amsmath}
\usepackage{amsthm}
\usepackage{braket}
\usepackage{indentfirst}
\usepackage{float}
\usepackage{multirow}
\usepackage[colorlinks]{hyperref}
\usepackage[dvipsnames]{xcolor}
\usepackage{xcolor}
\usepackage{booktabs}

\renewcommand\vec{\bm}

\definecolor{lgreen}{RGB}{0,180,0}
\definecolor{aqua}{RGB}{69,139,116}
\definecolor{oran}{RGB}{255,182,0}

\usepackage{subfigure}
\usepackage{algorithm}
\usepackage{algorithmicx}
\usepackage{algpseudocode}
\usepackage{amsmath}
\usepackage{verbatim}
\usepackage{makecell}
\usepackage[normalem]{ulem}

\begin{document}

\title{Active Learning on a Programmable Photonic Quantum Processor}

\author{Chen Ding}
\thanks{These two authors contributed equally.}
\affiliation{Henan Key Laboratory of Quantum Information and Cryptography, Zhengzhou, Henan 450000, China}

\author{Xiao-Yue Xu}
\thanks{These two authors contributed equally.}
\affiliation{Henan Key Laboratory of Quantum Information and Cryptography, Zhengzhou, Henan 450000, China}

\author{Yun-Fei Niu}
\affiliation{Henan Key Laboratory of Quantum Information and Cryptography, Zhengzhou, Henan 450000, China}
\author{Shuo Zhang}
\affiliation{Henan Key Laboratory of Quantum Information and Cryptography, Zhengzhou, Henan 450000, China}
\author{Wan-Su Bao}
\email{bws@qiclab.cn}
\affiliation{Henan Key Laboratory of Quantum Information and Cryptography, Zhengzhou, Henan 450000, China}
\author{He-Liang Huang}
\email{quanhhl@ustc.edu.cn}
\affiliation{Henan Key Laboratory of Quantum Information and Cryptography, Zhengzhou, Henan 450000, China}
\affiliation{Hefei National Research Center for Physical Sciences at the Microscale and School of Physical Sciences, University of Science and Technology of China, Hefei 230026, China}
\affiliation{Shanghai Research Center for Quantum Science and CAS Center for Excellence in Quantum Information and Quantum Physics, University of Science and Technology of China, Shanghai 201315, China}
\affiliation{Hefei National Laboratory, University of Science and Technology of China, Hefei 230088, China}

\date{\today}

\begin{abstract}
Training a quantum machine learning model generally requires a large labeled dataset, which incurs high labeling and computational costs. To reduce such costs, a selective training strategy, called active learning (AL), chooses only a subset of the original dataset to learn while maintaining the trained model's performance. Here, we design and implement two AL-enpowered variational quantum classifiers, to investigate the potential applications and effectiveness of AL in quantum machine learning. Firstly, we build a programmable free-space photonic quantum processor, which enables the programmed implementation of various hybrid quantum-classical computing algorithms. Then, we code the designed variational quantum classifier with AL into the quantum processor, and execute comparative tests for the classifiers with and without the AL strategy.
The results validate the great advantage of AL in quantum machine learning, as it saves at most $85\%$ labeling efforts and $91.6\%$ percent computational efforts compared to the training without AL on a data classification task.
Our results inspire AL's further applications in large-scale quantum machine learning to drastically reduce training data and speed up training, underpinning the exploration of practical quantum advantages in quantum physics or real-world applications.


\end{abstract}

\maketitle
\section{Introduction}

The fastly developing quantum machine learning algorithms could help solve numerous physical and real-world data analysis issues~\cite{biamonte2017quantum,kandala2017hardware,havlivcek2019supervised,schuld2019quantum,mcardle2020quantum,google2020hartree,huang2021experimental,liu2021hybrid,gong2022quantum}, and their potential computational advantages over classical counterparts have been investigated and even proven in a variety of theoretical and experimental works~\cite{zhou2020quantum,abbas2021power,gentini2020noise-resilient,bittel2021training,zheng2021speeding,golden2022evidence,saggio2021experimental,huang2022quantum, liu2021rigorous,yang2021provable,huang2021information, huang2018demonstration,ding2021noise,huang2020superconducting}. Despite the fact that the future subversion of quantum machine learning applications is rapidly approaching, there are still issues that need to be resolved for large-scale practical applications. One of the remaining problems is the high cost of labeling data items for supervised learning applications~\cite{morisio2020product,xu2022cross,wang2021want}. Since the labeling task requires finding a certain amount of answers to the question-to-solve, such work usually involves inefficient data collection or large human labour, especially for quantum machine learning models.
In applications like training quantum antoencoder~\cite{romero2017quantum,bondarenko2020quantum}, learning quantum dynamics~\cite{gong2022quantum, luchnikov2020machine} and solving real-world problems with powerful variational quantum models~\cite{huang2021experimental,liu2021hybrid,liu2021rigorous,endo2020variational,schuld2020circuit,cerezo2021variational}, preparing quantum states and labeling quantum state or quantum processes usually consumes huge physical resources. This drawback severely hurdles the further development of their practical applications. 

Active learning (AL)~\cite{baldridge2004active,settles2009active,huang2014active} is a effective method to resolve the above problem. Instead of acquiring all the labeled data to do the training job, AL strategies enable the model to query only a subset of the original data, while keeping the capability to faithfully extract the most critical features from the whole dataset. To achieve this goal, the model analyzes the unlabeled data pool and chooses only the representative data items to learn. Besides saving the labeling cost, active learning also accelerates the training process and may acquire computational advantage. Recent work has demonstrated its feasibility in quantum-related classical machine learning applications like retrieving quantum information~\cite{ding2020retrieving,dutt2021active,ding2022active} and create new quantum experiments~\cite{melnikov2018active}. However, its potential applications and practicality for quantum machine learning remains unknown.

In this paper, we implement two AL-enpowered variational quantum classifiers on a programmable photonic quantum processor. The whole system is built by high-quality single-photon sources, and circuit training and measurement devices connected to a control server, enabling fully programmed and high-precision manipulation (the average fidelity of the programmable single-qubit gate is $99.88\%$) of the expectation measurement, gradient evaluation and circuit training in the quantum processor. 
With the AL strategies coded into the program, we apply them to the variational quantum classifier and the nonlinearity-enhanced variational quantum classifier. 
Our comparative results confirm the efficacy of AL in reducing the cost of quantum machine learning, and its robust adaptability to different quantum machine learning models.
Furthermore, our techniques also demonstrate a path towards optical implementation of various types of hybrid quantum-classical algorithms in free-space photonic quantum processors.

\section{active learning}

\begin{figure}[t]
\begin{center}
\includegraphics[width=\linewidth]{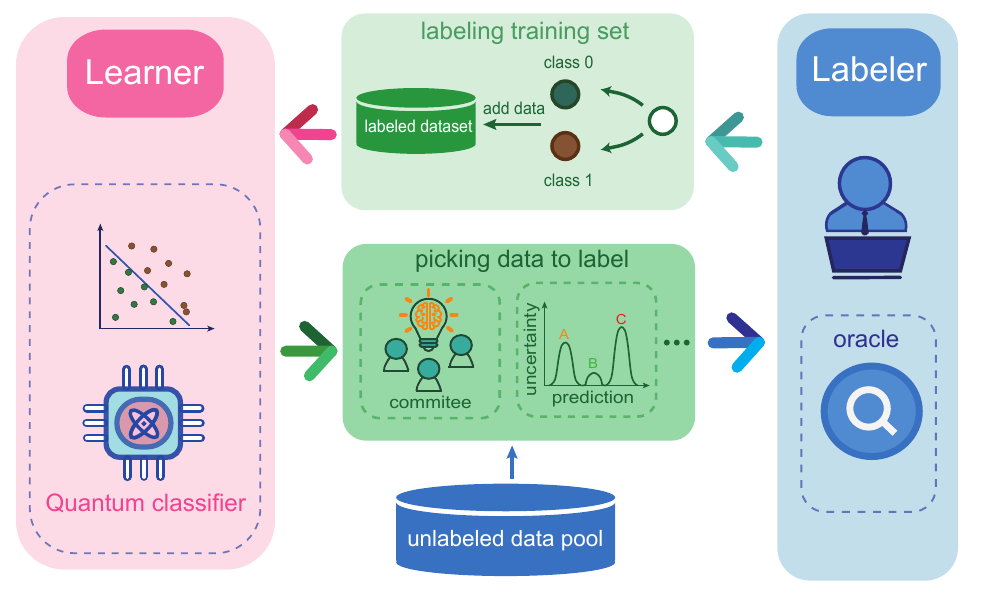}
\end{center}
\caption{\textbf{The schematic diagram of active learning.} From the unlabeled data pool, the learner sequentially selects a most representative data item (by evaluating the data uncertainty or consulting a committee of machine learning models) to query with the labeler. The labeler answers each query with the true label. The learner then trains the model for a short time period, and continues to select a new data item to learn until convergence.}
\label{fig-alg}
\end{figure}

Active learning is a type of strategies that applies selective technique to reduce the training set size, while preserving the performance of the original machine learning algorithm. Generally, we can view the training process of active learning as the cooperation work of a learner and a labeler, as shown in Fig.~\ref{fig-alg}. The learner is familiar with the model, the dataset, but unaware of their labels. Meanwhile, the labeler owns a labeling oracle and knows nothing else. The training works as follows:

1. In the beginning, the labeler sends a few labeled data items to the learner. The learner builds a prototype of the model using the given data as the training set.

2. By a certain selecting strategy, the learner selects one data item from the unlabeled data pool to query the labeler, then adds the labeled item to the training set.

3. The learner trains the model with the update labeled training set.

Repeat Steps 2 and 3 until convergence. To ensure effectiveness, the most representative data items could be extracted by some advanced selecting strategy, including uncertainty sampling (USAMP)~\cite{settles2008active,lewis1994sequential,scheffer2001active}, Query-by-Committee (QBC)~\cite{seung1992query}, evaluating expected model change~\cite{settles2007advances}, error reduction~\cite{roy2001toward}, etc. Those methods assess the informativeness of data items from different perspectives. Among them, two most typical techniques~\cite{ding2020retrieving}, USAMP and QBC, are chosen for our AL experiments and thus will be briefly introduced below.

Denote the training set as $\{x_i\},i=1,...,m$, their corresponding labels as $y_i$ in set $\mathcal{L}$. The USAMP method evaluates each data item's uncertainty for model $\theta$ as 
\begin{equation}
U(x_i)=-\underset{y\in \mathcal{L}} {\text{max}} P(\theta(x_i)=y),
\end{equation}
and select the data item with most uncertainty to query the labeler. 

The QBC method appoints a committee of several machine learning models, noted as $\theta_1,...,\theta_C$. For each data item, the committee members first guess its label separately, and then take a vote. The final selection goes to the most divergent item with the highest vote entropy
\begin{equation}
E(x_i) =-\sum_{j}\frac{V(y_j|x_i)}{C} \cdot \ln \frac{V(y_j|x_i)}{C},
\end{equation}
where $V(y_j|x_i)=\sum^C_{k=1}\delta_{\theta_k(x_i),y_j}$ denotes to the number of votes for label $y_i$ in the committee. 

\section{Experiment $\&$ Results}

In this section, we will implement the AL experiments on a programmable free-space photonic quantum processor (see Fig.~\ref{fig-experiment}). The experimental setup consists of two main modules, data encoding and variational quantum circuits (VQCs), both of which are programmable. Two types of VQCs with different capabilities are implemented in our experiments for the classification task. The results show that AL strategy has strong adaptability in different machine leaning models, and can greatly reduce the amount of data and computational cost.

\subsection{Data Encoding}\label{sec:data-encode}

\begin{figure*}
\begin{center}
\includegraphics[width=\linewidth]{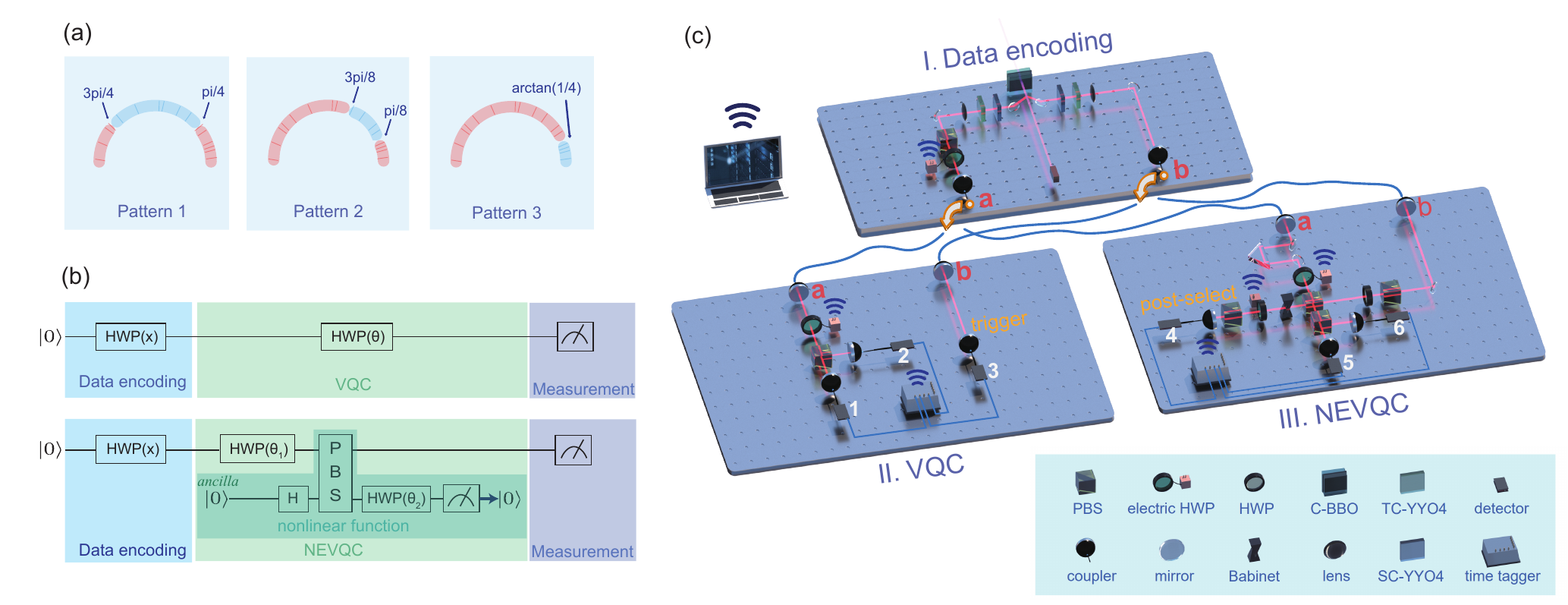}
\end{center}
\caption{\textbf{The experiment design and implementation.} (a) Three patterns of dividing the data items on an arc, designed to variate the difficulty of the classification task for variational quantum classifier. (b) The circuit of our variational quantum classifier (VQC) and the nonlinearity-enhanced variational quantum classifier (NEVQC). The nonlinearity, enhancing the performance of VQC on the data dividing patterns, is introduced by an entangling operations followed by postselecting the ancilla photon as $\ket{0}$. (c) The experimental setups on the photonic quantum processor. The implementations of data encoding (introduced in Sec.~\ref{sec:data-encode}), VQC (Sec.~\ref{sec:vqc}), NEVQC (Sec.~\ref{sec:nevqc}) are placed on three boards, connected through optical filbers. The photon detecters (noted in white numbers as 1-6) and electric rotation stages are also connected to the classical computer, automating the process of expectation measurement, gradient evaluation, and circuit training. The data photon and ancilla photon are noted as red letters as $a$ and $b$. On these boards, all photons are spectrally filtered with 3 nm bandwidth filters. Devices: C-BBO, sandwich-like BBO + HWP + BBO combination; QWP, quarter-wave plat; SC-YVO4, YVO4 crystal for spatial compensation; TC-YVO4, YVO4 crystal for temporal compensation; Babinet, Soleil-Babinet compensators.}
\label{fig-experiment}
\end{figure*}

As shown in Fig.~\ref{fig-experiment}(a), normalized vectors $(\cos x_i,\sin x_i)$ distributed on an arc of length $\pi$ are taken as the data for classification. These vectors are divided into two classes, labeling as $y_i=\pm 1$ (marked red and blue, respectively). To variate the difficulty of the classification, three patterns of dividing these vectors are designed.

To encode these data into quantum states with high fidelity, we first build a single-qubit photon source.
As shown on the Board I in Fig.~\ref{fig-experiment}(c), laser pulses with a central wavelength of 390 nm, pulse duration of 150 fs, and repetition rate of 80 MHz pass through a half-wave plate (HWP) sandwiched by two $\beta$-barium borate (BBO) crystals. By the spontaneous parametric down-conversion (SPDC) process, entangled photon pairs of visibility more than $100:1$ are produced on the two sides.
The two photons are in an entangled state $(\ket{HV}+\ket{VH})/\sqrt{2}$, where $\ket{H}$ represents the horizontal polarization and $\ket{V}$ represents the vertical polarization. We also note $\ket{H}$ as $\ket{0}$, and $\ket{V}$ as $\ket{1}$.

Then, a polarizing beam splitter (PBS) on one side postselects the photons as $\ket{HV}$, which disentangled them. 
We denote the photon of horizontal polarization as the ``data photon'', and the other photon as the ``ancilla photon''.
A single-qubit gate formed by a HWP at angle $x/2$ (denote by $\text{HWP}(x)$) encodes the data to the photon as
\begin{equation}\label{eqn:encode}
\begin{pmatrix}
\cos x &\sin x\\
\sin x&-\cos x
\end{pmatrix}\ket{H}=\cos x\ket{H}+\sin x\ket{V}.
\end{equation}

\begin{figure}
\begin{center}
\includegraphics[width=\linewidth]{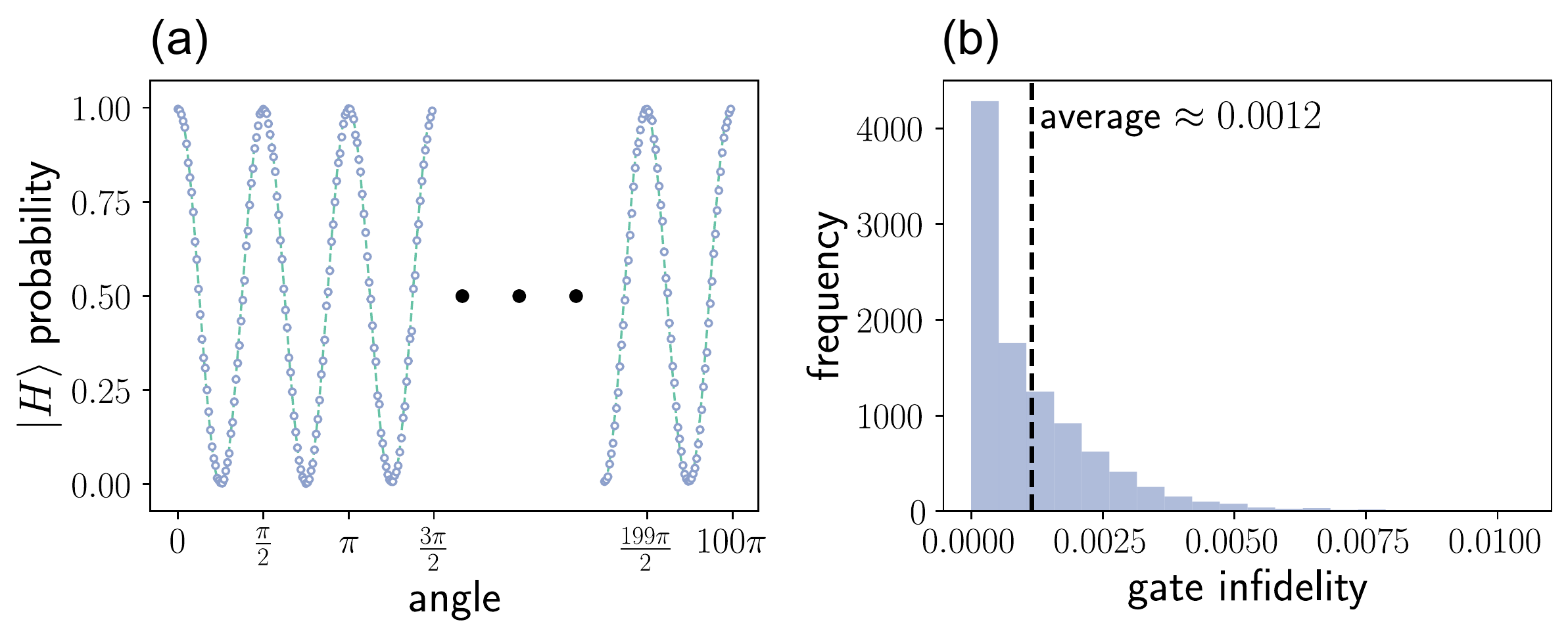}
\end{center}
\caption{\textbf{The calibration of HWP during long-term automatic rotation.} (a) The response curve of HWP during 10000 step rotation. We constantly rotate the HWP from angle 0 to $100\pi$ with step size $0.01\pi$. A single photon of horizontal polarization state passes through the HWP and is measured for 0.1 second after each rotation step. The dotted line shows the theoretical $\ket{H}$ probability in the photon, while the circles show the experimental measurement results, with around 2000 shots. We abbreviate the middle segment of the response curve for clarity of the result. (b) The distribution of calibrated gate infidelities of the HWP during the long-term rotation. The average infidelity is around $0.12\%$. Our calibration results show a high accuracy and stability of the single-qubit gate formed by HWP during long-term automatic rotation.}
\label{fig-cali}
\end{figure}

The whole dataset $\{(\cos x_i,\sin x_i)\}_{i=1,...,m}$ is encoded to the classifier successively. In a single epoch, each angle $x_i$ is visited by the HWP for a short duration, to evaluate a separate loss function or gradient. Then a classical computer will sum all the terms together and yield the parameters-to-be in the next epoch. 
Since the size of dataset is generally large in machine learning, a heavy HWP rotation work is brought. For instance, evaluating the loss function for 1000 epoches with dataset size of 100 will need about $10^5$ rotations of distance $500\pi$ on the encoding HWP, making manual experiment impossible, not to mention the subsequent low operation accuracy.  

To overcome this problem and accurately encode large datasets, we automate the rotation of the HWP by installing it into electric rotation stage and leave the large rotation work to a classical computer controlling the stage. We calibrate the response curve of the HWP to investigate their fidelity and stability during long-term electronically controlled rotations. The results in Fig.~\ref{fig-cali} show the HWP's response is in good agreement with the theoretical estimation during long-term automatic rotation, with the average fidelity estimated by the squared statistical overlap~\cite{fuchs1996distinguishability}
\begin{equation*}
\bar{F}=\frac{1}{10000}\sum{1-\left(\sqrt{p_iq_i}+\sqrt{(1-p_i)(1-q_i)}\right)^2},
\end{equation*}
as $99.88\%$ ($p_i,q_i$ are the theoretical and experimental $\ket{H}$ probability in the outcome photon, respectively, during the 10000 step rotation.). Our high-performance technique makes possible the training of large quantum machine learning models, as we also apply it to the iterations of ansatz parameters in model training.

\subsection{Variational Quantum Classifier (VQC)}\label{sec:vqc}

\subsubsection{Ansatz}

For classifying the one-qubit data, we employ single-qubit parameterized quantum circuit consisting of a trainable single-qubit gate formed by a HWP, denoted by $\text{HWP}{(\theta)}$, as the trainable ansatz, as shown on the top panel in Fig.~\ref{fig-experiment}(b). The variational ansatz applys on the input states and analyzes the encoded information. The output states are all measured in Pauli-Z basis. The classification rule for state $\ket{x}$ is set as
\begin{align*}
\braket{Z(x,\theta)}&>0\rightarrow 1\\
\braket{Z(x,\theta)}&<0\rightarrow -1,
\end{align*}
in which $V(\cdot)$ is the parameterized quantum circuit, $\braket{Z(x_i,\theta)}=\braket{x_i|V^\dagger(\vec{\theta})ZV(\vec{\theta})|x_i}$ is the measured expectation.
The circuit parameter $\theta$ is optimized to minimize the mean squared error (MSE) loss function on the given dataset
\begin{align}\label{eqn:mse}
C(\vec{\theta})&=\sum_i (\braket{Z(x_i,\theta)}-y_i)^2.
\end{align}

Generally, the optimization method is gradient descent, in which the cost function gradient 
\begin{align}\label{eqn:gradient-original}
\frac{\text{d} C(\theta)}{\text{d} \theta}=\sum_i\left(\braket{Z(x_i,\theta)}-y_i\right)\frac{\text{d} \braket{Z(x_i,\theta)}}{\text{d} \theta},
\end{align}
is evaluated by parameter shift rule~\cite{schuld2019evaluating}
\begin{equation}\label{eqn:psr}
\frac{\text{d} \braket{Z(x_i,\theta)}}{\text{d} \theta} =\frac{1}{2}\left(\braket{Z(x_i,\theta+\frac{\pi}{4})}-\braket{Z(x_i,\theta-\frac{\pi}{4})}\right).
\end{equation}

\subsubsection{Experimental Implementation}

Our implementation of the circuit is shown on the Board II in Fig.~\ref{fig-experiment}(c), where the photon-a is used to perform the data encoding and variational quantum classifier. The measurement of the data photon is achieved by a PBS and two detectors on the transmission and reflection direction. The expectation is then calculated as
\begin{equation}\label{eqn:measurement}
\braket{Z}=\frac{N_{13}-N_{23}}{{N_{13}}+N_{23}},
\end{equation} 
in which $N_{13},N_{23}$ are the number of coincidence events among Detector 1-3 in 0.1 second. In Equation~(\ref{eqn:measurement}), the detection signal of ancilla photon (photon-b) is used as the trigger for the detection of the data photon. Partial higher-order events (e.g., double-pair emission) during SPDC, such as the noise events $N_{123}$, are automatically subtracted to improve the accuracy of the experiment.~\cite{wang2016experimental}. For each $x_i,\theta$, the overall number of measurement shots is around 2000.


We connect all the detectors to a classical computer that automatically evaluates Equation~(\ref{eqn:mse},\ref{eqn:gradient-original},\ref{eqn:measurement}) with the detection raw data. Since the HWPs are installed on electric rotation stage, we let the classical computer directs the iteration of parameter $\theta$ with the evaluated gradient and Adam optimizer~\cite{kingma2015Adam}, which forms a closed control loop. 

In each epoch, according to Equation~(\ref{eqn:gradient-original},\ref{eqn:psr}), expectations $\braket{Z(x_i,\theta\pm\pi/4)}$ and $\braket{Z(x_i,\theta)}$ need to be evaluated, which means the HWPs need to rotate to the angles $(x_i,\theta)$ and $(x_i,\theta\pm\pi/4)$. Since the HWPs can only be continously rotating, the total rotation distance (also the time consumption, given a constant rotating speed) during the training crucially depends on the rotation sequence of $\text{HWP}{(\theta)}$ and $\text{HWP}{(x)}$, presenting as the route passing through these angle positions. Though finding a shortest route passing the discrete points in the space (the travelling salesman problem), is generally hard to solve~\cite{drexl2015survey}, we develop following two techniques to speedup the training process and subsequently improve the computing accuracy.

Firstly, from the design of the single-qubit parameterized quantum circuit, it is easy to find
\begin{equation}\label{eqn:1q-property}
\braket{Z(x_i,\theta+\frac{\pi}{4})}+\braket{Z(x_i,\theta-\frac{\pi}{4})}=1.
\end{equation}

Then Equation~(\ref{eqn:gradient-original}) can be simplified as
\begin{align}\label{eqn:gradient}
\frac{\text{d} C(\theta)}{\text{d} \theta}=\sum_i(\braket{Z(x_i,\theta)}-y_i)(2\braket{Z(x_i,\theta+\frac{\pi}{4})}-1),
\end{align}
which reduces the required visiting positions to only $(x_i,\theta)$ and $(x_i,\theta+\pi/4)$, saving both the rotation steps and the rotation distance, which further reduces the cumulative error during the whole epoch.

\begin{figure}
\begin{center}
\includegraphics[width=\linewidth]{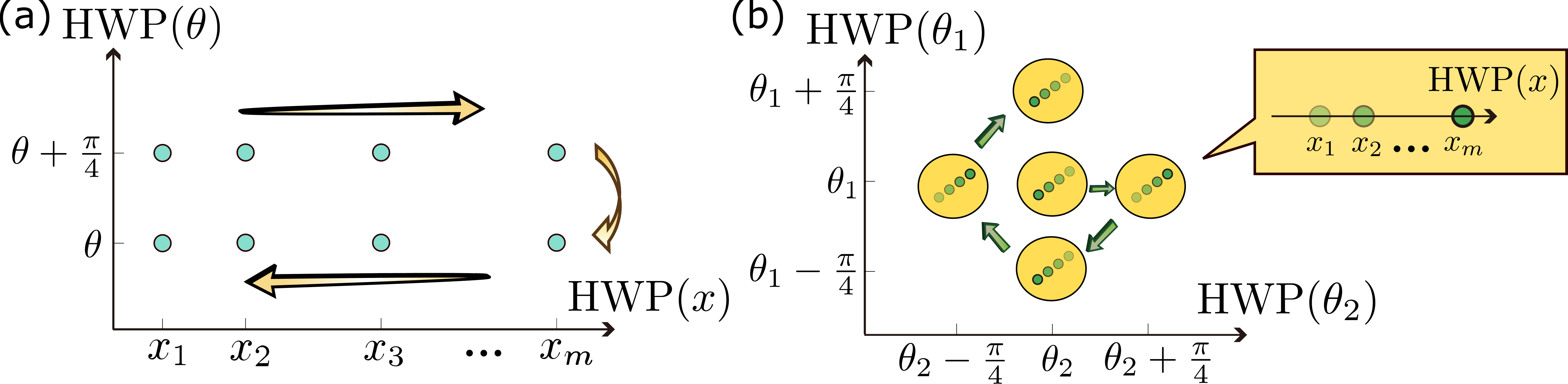}
\end{center}
\caption{\textbf{The optimized rotation sequences for the two variational quantum classifiers.} (a) The rotation squence of VQC. The points show the required visiting positions $(x_i,\theta), (x_i,\theta+\pi/4)$ in one iteration. The arrows show the rotation route. We first rotate $\text{HWP}{(\theta)}$ to $\theta+\pi/4$ and let $\text{HWP}(x)$ rotates from $x_1$ to $x_m$. Then we rotate $\text{HWP}(\theta$) to $\theta$ and let $\text{HWP}(x)$ rotates back from $x_m$ to $x_1$. (b) The rotation squence of NEVQC. The positions $(x_i,\theta_2,\theta_1), (x_i,\theta_2\pm\pi/4,\theta_1), (x_i,\theta_2,\theta_1\pm\pi/4)$ are distributed in three-dimensional space. Since the number of data items $x_i$ is generally large, we divided the positions into five groups. Each group contains a set of positions with same $\theta$s. We let the classifier preferentially travels in the $x_i$ direction, which is illustrated as the color fading in each group.}
\label{fig-journey}
\end{figure}

\begin{figure*}[t]
    \begin{center}
    \includegraphics[width=\linewidth]{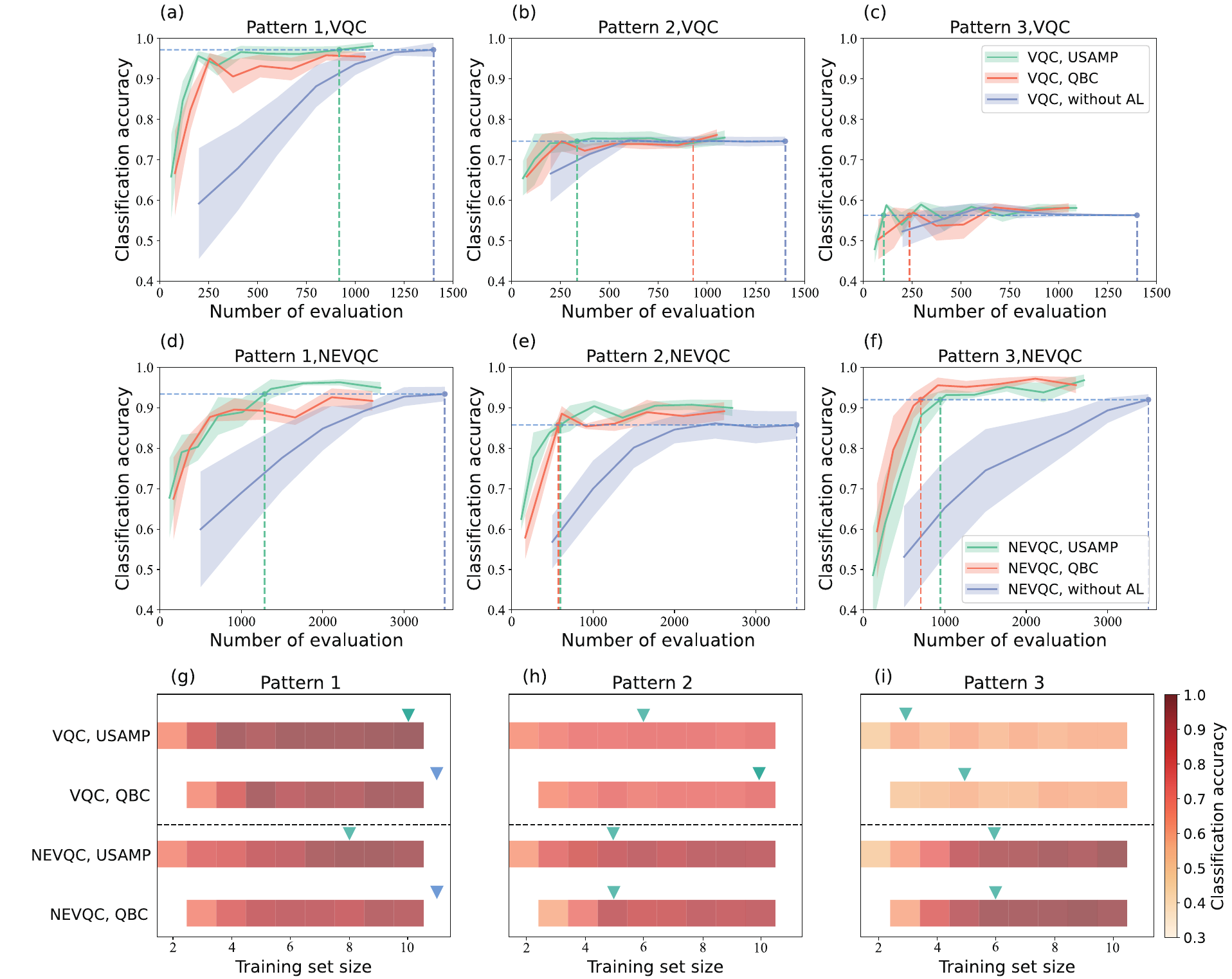}
    \end{center}
    \caption{\textbf{The performance of VQC, NEVQC with and without AL strategies during the whole training process.} (a-f) The classification accuracies of VQC/NEVQC on the testing set with the varying number of expectation evaluations. Each line shows an average result over 4 indentical independent experiments and the bands show the standard deviation. The blue dashed lines mark the classification accuracies of VQC (without AL) at convergence and the corresponding numbers of evaluations. The green and orange dashed lines show the the required numbers of evaluations of VQC (with USAMP and QBC, respectively) that achieve the same classification accuracies as VQC without AL. (g-i) The classification accuracies of VQC/NEVQC (with AL training strategies) on the testing set with varying training set size. The color of each horizontal bar shows the average classification accuracy with the corresponding training set size marked on the $x$-axis. The green triangles on the bars show the required training set size to achieve the same classification accuracies as VQC without AL. The blue triangles suggests cases that the final classification accuracies of VQC/NEVQC with QBC strategy and the 10-data-item training set cannot match the final classification accuracies of VQC/NEVQC without AL (but already very close). Among these subfigures, each column corresponds to each dataset dividing patterns, shown in Fig.~\ref{fig-experiment}(a). Tab.~\ref{ratios} shows the corresponding labeling and computational cost ratios for classifiers with and without AL.
    }
    \label{fig-out}
    \end{figure*}

Secondly, we note the positions we need to pass through are distributed on two parallel lines, where the number of $x_i$ (also the size of training set) is generally large. To avoid the frequent round trip of $\text{HWP}{(\theta)}$, we let the program travels along the $x_i$ direction with priority, which forms a U-shaped route demonstrated in Fig.~\ref{fig-journey}(a).

\subsubsection{Results}

We first train the classifier without AL for 35 steps with 20 labeled data vectors of three dividing patterns. Then we adopt the USAMP and the QBC strategies  to the classifier respectively to observe the performance improvement of AL. We let the classifier actively selects 10 samples to learn from an unlabeled data pool of 20 data items, starting from the given 2 (for USAMP) or 3 (for QBC) data items with different labels. For each sample, the classifier trains itself for 10 epoches  with the data it has. The members of the committee for QBC strategy are Support Vector Classifier (SVC)~\cite{cortes1995support, Keerthi2003asymptotic} with the Radial Basis Function (RBF) kernel, k-nearest neighbor classification~\cite{cover1967nearest, bailey1978anote} with $k=3$, Linear Discriminant Analysis algorithm~\cite{fisher1936theuse, carl1975discriminant}, and decision tree~\cite{quinlan1986induction, safavian1991asurvey} with max depth 7. These models are detailedly introduced in Supplemental Material. Their hyperparameters are pre-optimized to fit our classification tasks here.
To monitor the classifiers' performance during the training process, we timely benchmark the classifier's performance. We do the test for 5 (for without AL) or 10 (for AL) epoches by a testing set of the corresponding training set pattern with 500 vectors. We also conduct corresponding numerical simulations (See results in Supplemental Material) to validate the experiment results.

\begin{table*}[t]
\centering
\caption{The labeling and computational cost ratios (AL/non-AL) in the training of VQC/NEVQC. The columns ``minimum'' and ``mean'' show the minimum and average ratios in each row. The mark $\times$ in the table represents the cases that the final classification accuracies of VQC/NEVQC with QBC strategy and the 10-data-item training set cannot match the final classification accuracies of VQC/NEVQC without AL (but already very close). The results generally show a sharp cut of labeling efforts and computational cost by the AL strategies.}\label{ratios}
\begin{tabular}{c|c||c|c||c|c||c|c||c|c} 
\hline
\multicolumn{2}{c||}{\multirow{2}{*}{Ratios($\%$)}} & \multicolumn{2}{c||}{Pattern 1} & \multicolumn{2}{c||}{Pattern 2} & \multicolumn{2}{c||}{Pattern 3} & \multirow{2}{*}{$\ $~~minimum~~$\ $} & \multirow{2}{*}{$\ $~~mean~~$\ $}  \\ 
\cline{3-8}
\multicolumn{2}{c||}{}                              & $\ $~~USAMP~~$\ $    & $\ $~~QBC~~$\ $                       & $\ $~~USAMP~~$\ $    & ~~QBC~~                        & $\ $~~USAMP~~$\ $   & $\ $~~QBC~~$\ $                         &                        &                        \\ 
\hline
\multirow{2}{*}{~~VQC~~}   & ~~labeling~~                   & 50   & $\times$                 & 30   & 50                       & 15  & 25                        & 15                     & 34                     \\ 
\cline{2-10}
&$\ $ ~~computation~~   $\ $             & 77.9 & $\times$                 & 29.6 & 75                       & 8.4 & 18.2                      & 8.4                    & 46.8                   \\ 
\hline
\multirow{2}{*}{~~NEVQC~~} & ~~labeling~~                   & 35   & $\times$                 & 25   & 25                       & 30  & 30                        & 25                     & 29                     \\ 
\cline{2-10}
&$\ $ ~~computation~~  $\ $              & 38.9 & $\times$                 & 20.4 & 17.6                     & 29  & 26.1                      & 17.6                   & 26.4                   \\
\hline
\end{tabular}
\end{table*}

The classification accuracies during the whole training process are shown in Fig.~\ref{fig-out}(a-c). The performance of training without AL, and the training with two active learning strategies are compared together with the same ``number of evaluations'', which is calculated as the cumulative times of expectation evaluation, directly corresponding to the realistic time consumption. 
The results suggest the AL training strategies generally acclerate the training as they achieves same classification accuracy with smaller number of evaluations than training without AL. We show the ratios of labeling and computational costs (AL/non-AL) in Tab.~\ref{ratios}. The labeling costs of the methods are the numbers of labeled data items for training, and the computational costs are calculated as their number of evaluations. From the table, we can see that the AL can achieve a maximum reduction of $91.6\%$ computing resources, compared with the classifier without AL, in the case of using the USAMP strategy for Patten 3. Figure~\ref{fig-out}(g-i) show the classification accuracies varying with training set size. We find the performance of both active learning strategies becomes better as they expand the training set size. In most cases, compared to the classifier without AL, the classifier with AL achieves the same classification accuracy (near convergence point) in the training set of less than 10 data items. As can be seen from Tab.~\ref{ratios}, the labeling efforts can be reduced by up to $85\%$, in the case of using the USAMP strategy for Patten 3.

The single-qubit classifier performs best on Pattern 1, while worst on Pattern 3, showing its endogenous classification ability, as its dividing boundaries are always vertical (see a rigorous proof in Supplemental Material). To further boost the performance, we strengthen the classifier by adding a nonlinear operation to the model.

\subsection{Nonlinearity-Enhanced Variational Quantum Classifier (NEVQC)}\label{sec:nevqc}

We introduce nonlinearity to the ansatz to enhance the power of our variational quantum classifier~
\cite{aaronson2005quantum,holmes2021nonlinear}, by introducing an ancilla qubit and some additional quantum operations. As shown on the bottom panel in Fig.~\ref{fig-experiment}(b) and the Board III in Fig.~\ref{fig-experiment}(c).
We first apply a Hadamard gate on the ancilla photon to prepare $(\ket{H}+\ket{V})/\sqrt{2}$. Then, we apply a PBS on the two photons and postselect the events where there is exactly one photon exiting each output of the PBS. 
Two parameterized single-qubit gates are inserted into the circuit for training the model. Finally, we postselect $\ket{0}$ events in the ancilla and measure the input qubit in the Z basis. The ancilla photon, an additional Hadamard gate and PBS operation, and post-selection of measurements, constitute the introduced nonlinearity. We note that such nonlinearity does not even require the occurance of interference on the PBS (See proof in Supplemental Material). Therefore, we do not need to intentionally balance the optical distance difference between the data photon and the ancilla photon, which is experiment friendly. We also conduct the interference version of the same experiment and show the results in Supplemental Material.   

The measured $Z$ expectation for photon-a is calculated as
\begin{equation}\label{eqn:measurement2}
\braket{Z}=\frac{N_{45}-N_{46}}{{N_{45}}+N_{46}},
\end{equation}
in which $N_{45},N_{46}$ are the number of coincidence events among Detector 4-6 in one second. The number of measurement shots is 5500.

The same experiment settings, including the cost function, the active learning strategies, the dataset configuration, are applied for NEVQC as for VQC. However, the automatic rotation scheme of the electrically driven HWP (trainable single-qubit gate) slightly different, since we now have three electrically driven HWPs ($\text{HWP}{(x)}$ for encoding the data and $\text{HWP}{(\theta_1)},\text{HWP}(\theta_2)$ for the two trainable single-qubit gates), thus the corresponding route is then in 3-dimensional space. Moreover, Equation~(\ref{eqn:1q-property}) no longer holds. The program needs to do the measurements with the angles as $(x_i,\theta_2,\theta_1), (x_i,\theta_2\pm\pi/4,\theta_1), (x_i,\theta_2,\theta_1\pm\pi/4)$ for the evaluation of derivatives $\partial C/\partial \theta_2$ and $\partial C/\partial \theta_1$.
Our strategy is yet similar. As shown in Fig.~\ref{fig-journey}(b), same as in the training of VQC, we let the program preferentially travel in the $x_i$ direction, and then seek the shortest path among the five parameter configurations ($\{(\theta_1,\theta_2),(\theta_1\pm\pi/4,\theta_2\pm\pi/4)\}$) in the left two dimension.

The classification accuracies during the whole training process in Fig.~\ref{fig-out}(d-f) show the advantage of NEVQC compared to VQC, as it generally achieves higher classification accuracies on the three patterns, which fully demonstrates the performance improvement after we introduce nonlinear operation. The efficacy of active learning is also validated on NEVQC, as USAMP and QBC achieves same classification accuracy in less time than the training without AL in most cases. As shown in Tab.~\ref{ratios}, the training with AL strategies require only $17.6\%$ number of evaluations to match the performance of training without AL in the best case.
Meanwhile, the active learning methods also enhance the upper bound of the model's performance, which may originate from their representive choice of data.
Fig.~\ref{fig-out}(g-i) shows the classification accuracies varying with training set size. The phenomena is similar with the ones of VQC. In the best case, training with AL strategies only require $25\%$ labeling efforts to match the performance of training without AL. 

\section{Conclusion}
We build a fully-programmable photonic quantum processor with high gate fidelities, which allows flexible execution of hybrid quantum-classical computing schemes. The AL strategies are coded as classical programs and deployed to our designed variational quantum classifiers on the quantum processor. Comparative experiments demonstrate a very positive effect of AL in the practical application of quantum machine learning, as it sharply cuts $66\%$ labeling requirements and $53\%$ computations on average, and $85\%$ labeling efforts and $91.6\%$ computations at most. 

Besides, the implementation of high accuracy operations, programmability, and speedup techniques during the training in the experiments make possible the implementations of various quantum machine learning algorithms in the free-space optical quantum system. And the experiment-friendly implementation of nonlinearity in our experiment also demonstrates the path towards designing powerful and practical quantum machine learning models. By introducing the large-scale interferometer similar to the state-of-the-art photonic quantum processor, \textit{JiuZhang}~\cite{zhong2020quantum,zhong2021phase}, it is possible to immediately form large-scale quantum machine learning applications.


\begin{acknowledgments}
H.-L. H. acknowledges support from the Youth Talent Lifting Project (Grant No. 2020-JCJQ-QT-030), National Natural Science Foundation of China (Grants No. 11905294), China Postdoctoral Science Foundation, and the Open Research Fund from State Key Laboratory of High Performance Computing of China (Grant No. 201901-01).
\end{acknowledgments}
\bibliography{b}

\begin{thebibliography}{62}%
\makeatletter
\providecommand \@ifxundefined [1]{%
 \@ifx{#1\undefined}
}%
\providecommand \@ifnum [1]{%
 \ifnum #1\expandafter \@firstoftwo
 \else \expandafter \@secondoftwo
 \fi
}%
\providecommand \@ifx [1]{%
 \ifx #1\expandafter \@firstoftwo
 \else \expandafter \@secondoftwo
 \fi
}%
\providecommand \natexlab [1]{#1}%
\providecommand \enquote  [1]{``#1''}%
\providecommand \bibnamefont  [1]{#1}%
\providecommand \bibfnamefont [1]{#1}%
\providecommand \citenamefont [1]{#1}%
\providecommand \href@noop [0]{\@secondoftwo}%
\providecommand \href [0]{\begingroup \@sanitize@url \@href}%
\providecommand \@href[1]{\@@startlink{#1}\@@href}%
\providecommand \@@href[1]{\endgroup#1\@@endlink}%
\providecommand \@sanitize@url [0]{\catcode `\\12\catcode `\$12\catcode
  `\&12\catcode `\#12\catcode `\^12\catcode `\_12\catcode `\%12\relax}%
\providecommand \@@startlink[1]{}%
\providecommand \@@endlink[0]{}%
\providecommand \url  [0]{\begingroup\@sanitize@url \@url }%
\providecommand \@url [1]{\endgroup\@href {#1}{\urlprefix }}%
\providecommand \urlprefix  [0]{URL }%
\providecommand \Eprint [0]{\href }%
\providecommand \doibase [0]{https://doi.org/}%
\providecommand \selectlanguage [0]{\@gobble}%
\providecommand \bibinfo  [0]{\@secondoftwo}%
\providecommand \bibfield  [0]{\@secondoftwo}%
\providecommand \translation [1]{[#1]}%
\providecommand \BibitemOpen [0]{}%
\providecommand \bibitemStop [0]{}%
\providecommand \bibitemNoStop [0]{.\EOS\space}%
\providecommand \EOS [0]{\spacefactor3000\relax}%
\providecommand \BibitemShut  [1]{\csname bibitem#1\endcsname}%
\let\auto@bib@innerbib\@empty
\bibitem [{\citenamefont {Biamonte}\ \emph {et~al.}(2017)\citenamefont
  {Biamonte}, \citenamefont {Wittek}, \citenamefont {Pancotti}, \citenamefont
  {Rebentrost}, \citenamefont {Wiebe},\ and\ \citenamefont
  {Lloyd}}]{biamonte2017quantum}%
  \BibitemOpen
  \bibfield  {author} {\bibinfo {author} {\bibfnamefont {J.}~\bibnamefont
  {Biamonte}}, \bibinfo {author} {\bibfnamefont {P.}~\bibnamefont {Wittek}},
  \bibinfo {author} {\bibfnamefont {N.}~\bibnamefont {Pancotti}}, \bibinfo
  {author} {\bibfnamefont {P.}~\bibnamefont {Rebentrost}}, \bibinfo {author}
  {\bibfnamefont {N.}~\bibnamefont {Wiebe}},\ and\ \bibinfo {author}
  {\bibfnamefont {S.}~\bibnamefont {Lloyd}},\ }\href
  {https://doi.org/10.1038/nature23474} {\bibfield  {journal} {\bibinfo
  {journal} {Nature}\ }\textbf {\bibinfo {volume} {549}},\ \bibinfo {pages}
  {195} (\bibinfo {year} {2017})}\BibitemShut {NoStop}%
\bibitem [{\citenamefont {Kandala}\ \emph {et~al.}(2017)\citenamefont
  {Kandala}, \citenamefont {Mezzacapo}, \citenamefont {Temme}, \citenamefont
  {Takita}, \citenamefont {Brink}, \citenamefont {Chow},\ and\ \citenamefont
  {Gambetta}}]{kandala2017hardware}%
  \BibitemOpen
  \bibfield  {author} {\bibinfo {author} {\bibfnamefont {A.}~\bibnamefont
  {Kandala}}, \bibinfo {author} {\bibfnamefont {A.}~\bibnamefont {Mezzacapo}},
  \bibinfo {author} {\bibfnamefont {K.}~\bibnamefont {Temme}}, \bibinfo
  {author} {\bibfnamefont {M.}~\bibnamefont {Takita}}, \bibinfo {author}
  {\bibfnamefont {M.}~\bibnamefont {Brink}}, \bibinfo {author} {\bibfnamefont
  {J.~M.}\ \bibnamefont {Chow}},\ and\ \bibinfo {author} {\bibfnamefont
  {J.~M.}\ \bibnamefont {Gambetta}},\ }\href
  {https://doi.org/10.1038/nature23879} {\bibfield  {journal} {\bibinfo
  {journal} {Nature}\ }\textbf {\bibinfo {volume} {549}},\ \bibinfo {pages}
  {242} (\bibinfo {year} {2017})}\BibitemShut {NoStop}%
\bibitem [{\citenamefont {Havlíček}\ \emph {et~al.}(2019)\citenamefont
  {Havlíček}, \citenamefont {Córcoles}, \citenamefont {Temme}, \citenamefont
  {Harrow}, \citenamefont {Kandala}, \citenamefont {Chow},\ and\ \citenamefont
  {Gambetta}}]{havlivcek2019supervised}%
  \BibitemOpen
  \bibfield  {author} {\bibinfo {author} {\bibfnamefont {V.}~\bibnamefont
  {Havlíček}}, \bibinfo {author} {\bibfnamefont {A.~D.}\ \bibnamefont
  {Córcoles}}, \bibinfo {author} {\bibfnamefont {K.}~\bibnamefont {Temme}},
  \bibinfo {author} {\bibfnamefont {A.~W.}\ \bibnamefont {Harrow}}, \bibinfo
  {author} {\bibfnamefont {A.}~\bibnamefont {Kandala}}, \bibinfo {author}
  {\bibfnamefont {J.~M.}\ \bibnamefont {Chow}},\ and\ \bibinfo {author}
  {\bibfnamefont {J.~M.}\ \bibnamefont {Gambetta}},\ }\href
  {https://doi.org/10.1038/s41586-019-0980-2} {\bibfield  {journal} {\bibinfo
  {journal} {Nature}\ }\textbf {\bibinfo {volume} {567}},\ \bibinfo {pages}
  {209} (\bibinfo {year} {2019})}\BibitemShut {NoStop}%
\bibitem [{\citenamefont {Schuld}\ and\ \citenamefont
  {Killoran}(2019)}]{schuld2019quantum}%
  \BibitemOpen
  \bibfield  {author} {\bibinfo {author} {\bibfnamefont {M.}~\bibnamefont
  {Schuld}}\ and\ \bibinfo {author} {\bibfnamefont {N.}~\bibnamefont
  {Killoran}},\ }\href {https://doi.org/10.1103/PhysRevLett.122.040504}
  {\bibfield  {journal} {\bibinfo  {journal} {Phys. Rev. Lett.}\ }\textbf
  {\bibinfo {volume} {122}},\ \bibinfo {pages} {040504} (\bibinfo {year}
  {2019})}\BibitemShut {NoStop}%
\bibitem [{\citenamefont {McArdle}\ \emph {et~al.}(2020)\citenamefont
  {McArdle}, \citenamefont {Endo}, \citenamefont {Aspuru-Guzik}, \citenamefont
  {Benjamin},\ and\ \citenamefont {Yuan}}]{mcardle2020quantum}%
  \BibitemOpen
  \bibfield  {author} {\bibinfo {author} {\bibfnamefont {S.}~\bibnamefont
  {McArdle}}, \bibinfo {author} {\bibfnamefont {S.}~\bibnamefont {Endo}},
  \bibinfo {author} {\bibfnamefont {A.}~\bibnamefont {Aspuru-Guzik}}, \bibinfo
  {author} {\bibfnamefont {S.~C.}\ \bibnamefont {Benjamin}},\ and\ \bibinfo
  {author} {\bibfnamefont {X.}~\bibnamefont {Yuan}},\ }\href
  {https://doi.org/10.1103/RevModPhys.92.015003} {\bibfield  {journal}
  {\bibinfo  {journal} {Rev. Mod. Phys.}\ }\textbf {\bibinfo {volume} {92}},\
  \bibinfo {pages} {015003} (\bibinfo {year} {2020})}\BibitemShut {NoStop}%
\bibitem [{\citenamefont {Quantum}\ \emph {et~al.}(2020)\citenamefont
  {Quantum}, \citenamefont {Collaborators*†}, \citenamefont {Arute},
  \citenamefont {Arya}, \citenamefont {Babbush}, \citenamefont {Bacon},
  \citenamefont {Bardin}, \citenamefont {Barends}, \citenamefont {Boixo},
  \citenamefont {Broughton}, \citenamefont {Buckley} \emph
  {et~al.}}]{google2020hartree}%
  \BibitemOpen
  \bibfield  {author} {\bibinfo {author} {\bibfnamefont {G.~A.}\ \bibnamefont
  {Quantum}}, \bibinfo {author} {\bibnamefont {Collaborators*†}}, \bibinfo
  {author} {\bibfnamefont {F.}~\bibnamefont {Arute}}, \bibinfo {author}
  {\bibfnamefont {K.}~\bibnamefont {Arya}}, \bibinfo {author} {\bibfnamefont
  {R.}~\bibnamefont {Babbush}}, \bibinfo {author} {\bibfnamefont
  {D.}~\bibnamefont {Bacon}}, \bibinfo {author} {\bibfnamefont {J.~C.}\
  \bibnamefont {Bardin}}, \bibinfo {author} {\bibfnamefont {R.}~\bibnamefont
  {Barends}}, \bibinfo {author} {\bibfnamefont {S.}~\bibnamefont {Boixo}},
  \bibinfo {author} {\bibfnamefont {M.}~\bibnamefont {Broughton}}, \bibinfo
  {author} {\bibfnamefont {B.~B.}\ \bibnamefont {Buckley}}, \emph {et~al.},\
  }\href {https://doi.org/10.1126/science.abb9811} {\bibfield  {journal}
  {\bibinfo  {journal} {Science}\ }\textbf {\bibinfo {volume} {369}},\ \bibinfo
  {pages} {1084} (\bibinfo {year} {2020})}\BibitemShut {NoStop}%
\bibitem [{\citenamefont {Huang}\ \emph
  {et~al.}(2021{\natexlab{a}})\citenamefont {Huang}, \citenamefont {Du},
  \citenamefont {Gong}, \citenamefont {Zhao}, \citenamefont {Wu}, \citenamefont
  {Wang}, \citenamefont {Li}, \citenamefont {Liang}, \citenamefont {Lin},
  \citenamefont {Xu} \emph {et~al.}}]{huang2021experimental}%
  \BibitemOpen
  \bibfield  {author} {\bibinfo {author} {\bibfnamefont {H.-L.}\ \bibnamefont
  {Huang}}, \bibinfo {author} {\bibfnamefont {Y.}~\bibnamefont {Du}}, \bibinfo
  {author} {\bibfnamefont {M.}~\bibnamefont {Gong}}, \bibinfo {author}
  {\bibfnamefont {Y.}~\bibnamefont {Zhao}}, \bibinfo {author} {\bibfnamefont
  {Y.}~\bibnamefont {Wu}}, \bibinfo {author} {\bibfnamefont {C.}~\bibnamefont
  {Wang}}, \bibinfo {author} {\bibfnamefont {S.}~\bibnamefont {Li}}, \bibinfo
  {author} {\bibfnamefont {F.}~\bibnamefont {Liang}}, \bibinfo {author}
  {\bibfnamefont {J.}~\bibnamefont {Lin}}, \bibinfo {author} {\bibfnamefont
  {Y.}~\bibnamefont {Xu}}, \emph {et~al.},\ }\href
  {https://doi.org/10.1103/PhysRevApplied.16.024051} {\bibfield  {journal}
  {\bibinfo  {journal} {Phys. Rev. Applied}\ }\textbf {\bibinfo {volume}
  {16}},\ \bibinfo {pages} {024051} (\bibinfo {year}
  {2021}{\natexlab{a}})}\BibitemShut {NoStop}%
\bibitem [{\citenamefont {Liu}\ \emph {et~al.}(2021{\natexlab{a}})\citenamefont
  {Liu}, \citenamefont {Lim}, \citenamefont {Wood}, \citenamefont {Huang},
  \citenamefont {Guo},\ and\ \citenamefont {Huang}}]{liu2021hybrid}%
  \BibitemOpen
  \bibfield  {author} {\bibinfo {author} {\bibfnamefont {J.}~\bibnamefont
  {Liu}}, \bibinfo {author} {\bibfnamefont {K.~H.}\ \bibnamefont {Lim}},
  \bibinfo {author} {\bibfnamefont {K.~L.}\ \bibnamefont {Wood}}, \bibinfo
  {author} {\bibfnamefont {W.}~\bibnamefont {Huang}}, \bibinfo {author}
  {\bibfnamefont {C.}~\bibnamefont {Guo}},\ and\ \bibinfo {author}
  {\bibfnamefont {H.-L.}\ \bibnamefont {Huang}},\ }\href
  {https://doi.org/10.1007/s11433-021-1734-3} {\bibfield  {journal} {\bibinfo
  {journal} {Sci. China Phys. Mech.}\ }\textbf {\bibinfo {volume} {64}},\
  \bibinfo {pages} {290311} (\bibinfo {year} {2021}{\natexlab{a}})}\BibitemShut
  {NoStop}%
\bibitem [{\citenamefont {Gong}\ \emph {et~al.}(2022)\citenamefont {Gong},
  \citenamefont {Huang}, \citenamefont {Wang}, \citenamefont {Guo},
  \citenamefont {Li}, \citenamefont {Wu}, \citenamefont {Zhu}, \citenamefont
  {Zhao}, \citenamefont {Guo}, \citenamefont {Qian} \emph
  {et~al.}}]{gong2022quantum}%
  \BibitemOpen
  \bibfield  {author} {\bibinfo {author} {\bibfnamefont {M.}~\bibnamefont
  {Gong}}, \bibinfo {author} {\bibfnamefont {H.-L.}\ \bibnamefont {Huang}},
  \bibinfo {author} {\bibfnamefont {S.}~\bibnamefont {Wang}}, \bibinfo {author}
  {\bibfnamefont {C.}~\bibnamefont {Guo}}, \bibinfo {author} {\bibfnamefont
  {S.}~\bibnamefont {Li}}, \bibinfo {author} {\bibfnamefont {Y.}~\bibnamefont
  {Wu}}, \bibinfo {author} {\bibfnamefont {Q.}~\bibnamefont {Zhu}}, \bibinfo
  {author} {\bibfnamefont {Y.}~\bibnamefont {Zhao}}, \bibinfo {author}
  {\bibfnamefont {S.}~\bibnamefont {Guo}}, \bibinfo {author} {\bibfnamefont
  {H.}~\bibnamefont {Qian}}, \emph {et~al.},\ }\bibfield  {journal} {\bibinfo
  {journal} {arXiv:2201.05957}\ }\href
  {https://doi.org/https://doi.org/10.48550/arXiv.2201.05957}
  {https://doi.org/10.48550/arXiv.2201.05957} (\bibinfo {year}
  {2022})\BibitemShut {NoStop}%
\bibitem [{\citenamefont {Zhou}\ \emph {et~al.}(2020)\citenamefont {Zhou},
  \citenamefont {Wang}, \citenamefont {Choi}, \citenamefont {Pichler},\ and\
  \citenamefont {Lukin}}]{zhou2020quantum}%
  \BibitemOpen
  \bibfield  {author} {\bibinfo {author} {\bibfnamefont {L.}~\bibnamefont
  {Zhou}}, \bibinfo {author} {\bibfnamefont {S.-T.}\ \bibnamefont {Wang}},
  \bibinfo {author} {\bibfnamefont {S.}~\bibnamefont {Choi}}, \bibinfo {author}
  {\bibfnamefont {H.}~\bibnamefont {Pichler}},\ and\ \bibinfo {author}
  {\bibfnamefont {M.~D.}\ \bibnamefont {Lukin}},\ }\href
  {https://doi.org/10.1103/PhysRevX.10.021067} {\bibfield  {journal} {\bibinfo
  {journal} {Phys. Rev. X}\ }\textbf {\bibinfo {volume} {10}},\ \bibinfo
  {pages} {021067} (\bibinfo {year} {2020})}\BibitemShut {NoStop}%
\bibitem [{\citenamefont {Abbas}\ \emph {et~al.}(2021)\citenamefont {Abbas},
  \citenamefont {Sutter}, \citenamefont {Zoufal}, \citenamefont {Lucchi},
  \citenamefont {Figalli},\ and\ \citenamefont {Woerner}}]{abbas2021power}%
  \BibitemOpen
  \bibfield  {author} {\bibinfo {author} {\bibfnamefont {A.}~\bibnamefont
  {Abbas}}, \bibinfo {author} {\bibfnamefont {D.}~\bibnamefont {Sutter}},
  \bibinfo {author} {\bibfnamefont {C.}~\bibnamefont {Zoufal}}, \bibinfo
  {author} {\bibfnamefont {A.}~\bibnamefont {Lucchi}}, \bibinfo {author}
  {\bibfnamefont {A.}~\bibnamefont {Figalli}},\ and\ \bibinfo {author}
  {\bibfnamefont {S.}~\bibnamefont {Woerner}},\ }\href
  {https://doi.org/https://doi.org/10.1038/s43588-021-00084-1} {\bibfield
  {journal} {\bibinfo  {journal} {Nat. Comput. Sci.}\ }\textbf {\bibinfo
  {volume} {1}},\ \bibinfo {pages} {403} (\bibinfo {year} {2021})}\BibitemShut
  {NoStop}%
\bibitem [{\citenamefont {Gentini}\ \emph {et~al.}(2020)\citenamefont
  {Gentini}, \citenamefont {Cuccoli}, \citenamefont {Pirandola}, \citenamefont
  {Verrucchi},\ and\ \citenamefont {Banchi}}]{gentini2020noise-resilient}%
  \BibitemOpen
  \bibfield  {author} {\bibinfo {author} {\bibfnamefont {L.}~\bibnamefont
  {Gentini}}, \bibinfo {author} {\bibfnamefont {A.}~\bibnamefont {Cuccoli}},
  \bibinfo {author} {\bibfnamefont {S.}~\bibnamefont {Pirandola}}, \bibinfo
  {author} {\bibfnamefont {P.}~\bibnamefont {Verrucchi}},\ and\ \bibinfo
  {author} {\bibfnamefont {L.}~\bibnamefont {Banchi}},\ }\href
  {https://doi.org/10.1103/PhysRevA.102.052414} {\bibfield  {journal} {\bibinfo
   {journal} {Phys. Rev. A}\ }\textbf {\bibinfo {volume} {102}},\ \bibinfo
  {pages} {052414} (\bibinfo {year} {2020})}\BibitemShut {NoStop}%
\bibitem [{\citenamefont {Bittel}\ and\ \citenamefont
  {Kliesch}(2021)}]{bittel2021training}%
  \BibitemOpen
  \bibfield  {author} {\bibinfo {author} {\bibfnamefont {L.}~\bibnamefont
  {Bittel}}\ and\ \bibinfo {author} {\bibfnamefont {M.}~\bibnamefont
  {Kliesch}},\ }\href {https://doi.org/10.1103/PhysRevLett.127.120502}
  {\bibfield  {journal} {\bibinfo  {journal} {Phys. Rev. Lett.}\ }\textbf
  {\bibinfo {volume} {127}},\ \bibinfo {pages} {120502} (\bibinfo {year}
  {2021})}\BibitemShut {NoStop}%
\bibitem [{\citenamefont {Zheng}\ \emph {et~al.}(2021)\citenamefont {Zheng},
  \citenamefont {Li}, \citenamefont {Liu}, \citenamefont {Strelchuk},\ and\
  \citenamefont {Kondor}}]{zheng2021speeding}%
  \BibitemOpen
  \bibfield  {author} {\bibinfo {author} {\bibfnamefont {H.}~\bibnamefont
  {Zheng}}, \bibinfo {author} {\bibfnamefont {Z.}~\bibnamefont {Li}}, \bibinfo
  {author} {\bibfnamefont {J.}~\bibnamefont {Liu}}, \bibinfo {author}
  {\bibfnamefont {S.}~\bibnamefont {Strelchuk}},\ and\ \bibinfo {author}
  {\bibfnamefont {R.}~\bibnamefont {Kondor}},\ }\bibfield  {journal} {\bibinfo
  {journal} {arXiv:2112.07611}\ }\href
  {https://doi.org/https://doi.org/10.48550/arXiv.2112.07611}
  {https://doi.org/10.48550/arXiv.2112.07611} (\bibinfo {year}
  {2021})\BibitemShut {NoStop}%
\bibitem [{\citenamefont {Golden}\ \emph {et~al.}(2022)\citenamefont {Golden},
  \citenamefont {B{\"a}rtschi}, \citenamefont {Eidenbenz},\ and\ \citenamefont
  {O'Malley}}]{golden2022evidence}%
  \BibitemOpen
  \bibfield  {author} {\bibinfo {author} {\bibfnamefont {J.}~\bibnamefont
  {Golden}}, \bibinfo {author} {\bibfnamefont {A.}~\bibnamefont
  {B{\"a}rtschi}}, \bibinfo {author} {\bibfnamefont {S.}~\bibnamefont
  {Eidenbenz}},\ and\ \bibinfo {author} {\bibfnamefont {D.}~\bibnamefont
  {O'Malley}},\ }\bibfield  {journal} {\bibinfo  {journal} {arXiv:2202.00648}\
  }\href {https://doi.org/https://doi.org/10.48550/arXiv.2202.00648}
  {https://doi.org/10.48550/arXiv.2202.00648} (\bibinfo {year}
  {2022})\BibitemShut {NoStop}%
\bibitem [{\citenamefont {Saggio}\ \emph {et~al.}(2021)\citenamefont {Saggio},
  \citenamefont {Asenbeck}, \citenamefont {Hamann}, \citenamefont
  {Str{\"o}mberg}, \citenamefont {Schiansky}, \citenamefont {Dunjko},
  \citenamefont {Friis}, \citenamefont {Harris}, \citenamefont {Hochberg},
  \citenamefont {Englund} \emph {et~al.}}]{saggio2021experimental}%
  \BibitemOpen
  \bibfield  {author} {\bibinfo {author} {\bibfnamefont {V.}~\bibnamefont
  {Saggio}}, \bibinfo {author} {\bibfnamefont {B.~E.}\ \bibnamefont
  {Asenbeck}}, \bibinfo {author} {\bibfnamefont {A.}~\bibnamefont {Hamann}},
  \bibinfo {author} {\bibfnamefont {T.}~\bibnamefont {Str{\"o}mberg}}, \bibinfo
  {author} {\bibfnamefont {P.}~\bibnamefont {Schiansky}}, \bibinfo {author}
  {\bibfnamefont {V.}~\bibnamefont {Dunjko}}, \bibinfo {author} {\bibfnamefont
  {N.}~\bibnamefont {Friis}}, \bibinfo {author} {\bibfnamefont {N.~C.}\
  \bibnamefont {Harris}}, \bibinfo {author} {\bibfnamefont {M.}~\bibnamefont
  {Hochberg}}, \bibinfo {author} {\bibfnamefont {D.}~\bibnamefont {Englund}},
  \emph {et~al.},\ }\href
  {https://doi.org/https://doi.org/10.1038/s41586-021-03242-7} {\bibfield
  {journal} {\bibinfo  {journal} {Nature}\ }\textbf {\bibinfo {volume} {591}},\
  \bibinfo {pages} {229} (\bibinfo {year} {2021})}\BibitemShut {NoStop}%
\bibitem [{\citenamefont {Huang}\ \emph {et~al.}(2022)\citenamefont {Huang},
  \citenamefont {Broughton}, \citenamefont {Cotler}, \citenamefont {Chen},
  \citenamefont {Li}, \citenamefont {Mohseni}, \citenamefont {Neven},
  \citenamefont {Babbush}, \citenamefont {Kueng}, \citenamefont {Preskill}
  \emph {et~al.}}]{huang2022quantum}%
  \BibitemOpen
  \bibfield  {author} {\bibinfo {author} {\bibfnamefont {H.-Y.}\ \bibnamefont
  {Huang}}, \bibinfo {author} {\bibfnamefont {M.}~\bibnamefont {Broughton}},
  \bibinfo {author} {\bibfnamefont {J.}~\bibnamefont {Cotler}}, \bibinfo
  {author} {\bibfnamefont {S.}~\bibnamefont {Chen}}, \bibinfo {author}
  {\bibfnamefont {J.}~\bibnamefont {Li}}, \bibinfo {author} {\bibfnamefont
  {M.}~\bibnamefont {Mohseni}}, \bibinfo {author} {\bibfnamefont
  {H.}~\bibnamefont {Neven}}, \bibinfo {author} {\bibfnamefont
  {R.}~\bibnamefont {Babbush}}, \bibinfo {author} {\bibfnamefont
  {R.}~\bibnamefont {Kueng}}, \bibinfo {author} {\bibfnamefont
  {J.}~\bibnamefont {Preskill}}, \emph {et~al.},\ }\href
  {https://doi.org/10.1126/science.abn7293} {\bibfield  {journal} {\bibinfo
  {journal} {Science}\ }\textbf {\bibinfo {volume} {376}},\ \bibinfo {pages}
  {1182} (\bibinfo {year} {2022})}\BibitemShut {NoStop}%
\bibitem [{\citenamefont {Liu}\ \emph {et~al.}(2021{\natexlab{b}})\citenamefont
  {Liu}, \citenamefont {Arunachalam},\ and\ \citenamefont
  {Temme}}]{liu2021rigorous}%
  \BibitemOpen
  \bibfield  {author} {\bibinfo {author} {\bibfnamefont {Y.}~\bibnamefont
  {Liu}}, \bibinfo {author} {\bibfnamefont {S.}~\bibnamefont {Arunachalam}},\
  and\ \bibinfo {author} {\bibfnamefont {K.}~\bibnamefont {Temme}},\ }\href
  {https://doi.org/https://doi.org/10.1038/s41567-021-01287-z} {\bibfield
  {journal} {\bibinfo  {journal} {Nat. Phys.}\ }\textbf {\bibinfo {volume}
  {17}},\ \bibinfo {pages} {1013} (\bibinfo {year}
  {2021}{\natexlab{b}})}\BibitemShut {NoStop}%
\bibitem [{\citenamefont {Yang}\ \emph {et~al.}(2021)\citenamefont {Yang},
  \citenamefont {Garner}, \citenamefont {Liu}, \citenamefont {Tischler},
  \citenamefont {Thompson}, \citenamefont {Yung}, \citenamefont {Gu},\ and\
  \citenamefont {Dahlsten}}]{yang2021provable}%
  \BibitemOpen
  \bibfield  {author} {\bibinfo {author} {\bibfnamefont {C.}~\bibnamefont
  {Yang}}, \bibinfo {author} {\bibfnamefont {A.}~\bibnamefont {Garner}},
  \bibinfo {author} {\bibfnamefont {F.}~\bibnamefont {Liu}}, \bibinfo {author}
  {\bibfnamefont {N.}~\bibnamefont {Tischler}}, \bibinfo {author}
  {\bibfnamefont {J.}~\bibnamefont {Thompson}}, \bibinfo {author}
  {\bibfnamefont {M.-H.}\ \bibnamefont {Yung}}, \bibinfo {author}
  {\bibfnamefont {M.}~\bibnamefont {Gu}},\ and\ \bibinfo {author}
  {\bibfnamefont {O.}~\bibnamefont {Dahlsten}},\ }\bibfield  {journal}
  {\bibinfo  {journal} {arXiv:2105.14434}\ }\href
  {https://doi.org/https://doi.org/10.48550/arXiv.2105.14434}
  {https://doi.org/10.48550/arXiv.2105.14434} (\bibinfo {year}
  {2021})\BibitemShut {NoStop}%
\bibitem [{\citenamefont {Huang}\ \emph
  {et~al.}(2021{\natexlab{b}})\citenamefont {Huang}, \citenamefont {Kueng},\
  and\ \citenamefont {Preskill}}]{huang2021information}%
  \BibitemOpen
  \bibfield  {author} {\bibinfo {author} {\bibfnamefont {H.-Y.}\ \bibnamefont
  {Huang}}, \bibinfo {author} {\bibfnamefont {R.}~\bibnamefont {Kueng}},\ and\
  \bibinfo {author} {\bibfnamefont {J.}~\bibnamefont {Preskill}},\ }\href
  {https://doi.org/10.1103/PhysRevLett.126.190505} {\bibfield  {journal}
  {\bibinfo  {journal} {Phys. Rev. Lett.}\ }\textbf {\bibinfo {volume} {126}},\
  \bibinfo {pages} {190505} (\bibinfo {year} {2021}{\natexlab{b}})}\BibitemShut
  {NoStop}%
\bibitem [{\citenamefont {Huang}\ \emph {et~al.}(2018)\citenamefont {Huang},
  \citenamefont {Wang}, \citenamefont {Rohde}, \citenamefont {Luo},
  \citenamefont {Zhao}, \citenamefont {Liu}, \citenamefont {Li}, \citenamefont
  {Liu}, \citenamefont {Lu},\ and\ \citenamefont
  {Pan}}]{huang2018demonstration}%
  \BibitemOpen
  \bibfield  {author} {\bibinfo {author} {\bibfnamefont {H.-L.}\ \bibnamefont
  {Huang}}, \bibinfo {author} {\bibfnamefont {X.-L.}\ \bibnamefont {Wang}},
  \bibinfo {author} {\bibfnamefont {P.~P.}\ \bibnamefont {Rohde}}, \bibinfo
  {author} {\bibfnamefont {Y.-H.}\ \bibnamefont {Luo}}, \bibinfo {author}
  {\bibfnamefont {Y.-W.}\ \bibnamefont {Zhao}}, \bibinfo {author}
  {\bibfnamefont {C.}~\bibnamefont {Liu}}, \bibinfo {author} {\bibfnamefont
  {L.}~\bibnamefont {Li}}, \bibinfo {author} {\bibfnamefont {N.-L.}\
  \bibnamefont {Liu}}, \bibinfo {author} {\bibfnamefont {C.-Y.}\ \bibnamefont
  {Lu}},\ and\ \bibinfo {author} {\bibfnamefont {J.-W.}\ \bibnamefont {Pan}},\
  }\href {https://doi.org/10.1364/OPTICA.5.000193} {\bibfield  {journal}
  {\bibinfo  {journal} {Optica}\ }\textbf {\bibinfo {volume} {5}},\ \bibinfo
  {pages} {193} (\bibinfo {year} {2018})}\BibitemShut {NoStop}%
\bibitem [{\citenamefont {Ding}\ \emph {et~al.}(2021)\citenamefont {Ding},
  \citenamefont {Niu}, \citenamefont {Bao},\ and\ \citenamefont
  {Huang}}]{ding2021noise}%
  \BibitemOpen
  \bibfield  {author} {\bibinfo {author} {\bibfnamefont {C.}~\bibnamefont
  {Ding}}, \bibinfo {author} {\bibfnamefont {Y.-F.}\ \bibnamefont {Niu}},
  \bibinfo {author} {\bibfnamefont {W.-S.}\ \bibnamefont {Bao}},\ and\ \bibinfo
  {author} {\bibfnamefont {H.-L.}\ \bibnamefont {Huang}},\ }\bibfield
  {journal} {\bibinfo  {journal} {arXiv:2109.06805}\ }\href
  {https://doi.org/https://doi.org/10.48550/arXiv.2109.06805}
  {https://doi.org/10.48550/arXiv.2109.06805} (\bibinfo {year}
  {2021})\BibitemShut {NoStop}%
\bibitem [{\citenamefont {Huang}\ \emph {et~al.}(2020)\citenamefont {Huang},
  \citenamefont {Wu}, \citenamefont {Fan},\ and\ \citenamefont
  {Zhu}}]{huang2020superconducting}%
  \BibitemOpen
  \bibfield  {author} {\bibinfo {author} {\bibfnamefont {H.-L.}\ \bibnamefont
  {Huang}}, \bibinfo {author} {\bibfnamefont {D.}~\bibnamefont {Wu}}, \bibinfo
  {author} {\bibfnamefont {D.}~\bibnamefont {Fan}},\ and\ \bibinfo {author}
  {\bibfnamefont {X.}~\bibnamefont {Zhu}},\ }\href
  {https://doi.org/https://doi.org/10.1007/s11432-020-2881-9} {\bibfield
  {journal} {\bibinfo  {journal} {Sci. China Inf. Sci.}\ }\textbf {\bibinfo
  {volume} {63}},\ \bibinfo {pages} {1} (\bibinfo {year} {2020})}\BibitemShut
  {NoStop}%
\bibitem [{\citenamefont {Morisio}\ \emph {et~al.}(2020)\citenamefont
  {Morisio}, \citenamefont {Torchiano},\ and\ \citenamefont
  {Jedlitschka}}]{morisio2020product}%
  \BibitemOpen
  \bibfield  {author} {\bibinfo {author} {\bibfnamefont {M.}~\bibnamefont
  {Morisio}}, \bibinfo {author} {\bibfnamefont {M.}~\bibnamefont {Torchiano}},\
  and\ \bibinfo {author} {\bibfnamefont {A.}~\bibnamefont {Jedlitschka}},\
  }\href {https://doi.org/10.1007/978-3-030-64148-1} {\emph {\bibinfo {title}
  {Product-Focused Software Process Improvement: 21st International Conference,
  PROFES 2020, Turin, Italy, November 25--27, 2020, Proceedings}}},\ Vol.\
  \bibinfo {volume} {12562}\ (\bibinfo  {publisher} {Springer Nature},\
  \bibinfo {year} {2020})\BibitemShut {NoStop}%
\bibitem [{\citenamefont {Xu}\ \emph {et~al.}(2022)\citenamefont {Xu},
  \citenamefont {Wei}, \citenamefont {Sun}, \citenamefont {Yang}, \citenamefont
  {Shen}, \citenamefont {Dai}, \citenamefont {Zhou},\ and\ \citenamefont
  {Lin}}]{xu2022cross}%
  \BibitemOpen
  \bibfield  {author} {\bibinfo {author} {\bibfnamefont {Y.}~\bibnamefont
  {Xu}}, \bibinfo {author} {\bibfnamefont {F.}~\bibnamefont {Wei}}, \bibinfo
  {author} {\bibfnamefont {X.}~\bibnamefont {Sun}}, \bibinfo {author}
  {\bibfnamefont {C.}~\bibnamefont {Yang}}, \bibinfo {author} {\bibfnamefont
  {Y.}~\bibnamefont {Shen}}, \bibinfo {author} {\bibfnamefont {B.}~\bibnamefont
  {Dai}}, \bibinfo {author} {\bibfnamefont {B.}~\bibnamefont {Zhou}},\ and\
  \bibinfo {author} {\bibfnamefont {S.}~\bibnamefont {Lin}},\ }in\ \href
  {https://doi.org/https://doi.org/10.48550/arXiv.2112.09690} {\emph {\bibinfo
  {booktitle} {Proceedings of the IEEE/CVF Conference on Computer Vision and
  Pattern Recognition}}}\ (\bibinfo {year} {2022})\ pp.\ \bibinfo {pages}
  {2959--2968}\BibitemShut {NoStop}%
\bibitem [{\citenamefont {Wang}\ \emph {et~al.}(2021)\citenamefont {Wang},
  \citenamefont {Liu}, \citenamefont {Xu}, \citenamefont {Zhu},\ and\
  \citenamefont {Zeng}}]{wang2021want}%
  \BibitemOpen
  \bibfield  {author} {\bibinfo {author} {\bibfnamefont {S.}~\bibnamefont
  {Wang}}, \bibinfo {author} {\bibfnamefont {Y.}~\bibnamefont {Liu}}, \bibinfo
  {author} {\bibfnamefont {Y.}~\bibnamefont {Xu}}, \bibinfo {author}
  {\bibfnamefont {C.}~\bibnamefont {Zhu}},\ and\ \bibinfo {author}
  {\bibfnamefont {M.}~\bibnamefont {Zeng}},\ }\bibfield  {journal} {\bibinfo
  {journal} {arXiv:2108.13487}\ }\href
  {https://doi.org/https://doi.org/10.48550/arXiv.2108.13487}
  {https://doi.org/10.48550/arXiv.2108.13487} (\bibinfo {year}
  {2021})\BibitemShut {NoStop}%
\bibitem [{\citenamefont {Romero}\ \emph {et~al.}(2017)\citenamefont {Romero},
  \citenamefont {Olson},\ and\ \citenamefont
  {Aspuru-Guzik}}]{romero2017quantum}%
  \BibitemOpen
  \bibfield  {author} {\bibinfo {author} {\bibfnamefont {J.}~\bibnamefont
  {Romero}}, \bibinfo {author} {\bibfnamefont {J.~P.}\ \bibnamefont {Olson}},\
  and\ \bibinfo {author} {\bibfnamefont {A.}~\bibnamefont {Aspuru-Guzik}},\
  }\href {https://doi.org/10.1088/2058-9565/aa8072} {\bibfield  {journal}
  {\bibinfo  {journal} {Quantum Sci. Technol.}\ }\textbf {\bibinfo {volume}
  {2}},\ \bibinfo {pages} {045001} (\bibinfo {year} {2017})}\BibitemShut
  {NoStop}%
\bibitem [{\citenamefont {Bondarenko}\ and\ \citenamefont
  {Feldmann}(2020)}]{bondarenko2020quantum}%
  \BibitemOpen
  \bibfield  {author} {\bibinfo {author} {\bibfnamefont {D.}~\bibnamefont
  {Bondarenko}}\ and\ \bibinfo {author} {\bibfnamefont {P.}~\bibnamefont
  {Feldmann}},\ }\href {https://doi.org/10.1103/PhysRevLett.124.130502}
  {\bibfield  {journal} {\bibinfo  {journal} {Phys. Rev. Lett.}\ }\textbf
  {\bibinfo {volume} {124}},\ \bibinfo {pages} {130502} (\bibinfo {year}
  {2020})}\BibitemShut {NoStop}%
\bibitem [{\citenamefont {Luchnikov}\ \emph {et~al.}(2020)\citenamefont
  {Luchnikov}, \citenamefont {Vintskevich}, \citenamefont {Grigoriev},\ and\
  \citenamefont {Filippov}}]{luchnikov2020machine}%
  \BibitemOpen
  \bibfield  {author} {\bibinfo {author} {\bibfnamefont {I.~A.}\ \bibnamefont
  {Luchnikov}}, \bibinfo {author} {\bibfnamefont {S.~V.}\ \bibnamefont
  {Vintskevich}}, \bibinfo {author} {\bibfnamefont {D.~A.}\ \bibnamefont
  {Grigoriev}},\ and\ \bibinfo {author} {\bibfnamefont {S.~N.}\ \bibnamefont
  {Filippov}},\ }\href {https://doi.org/10.1103/PhysRevLett.124.140502}
  {\bibfield  {journal} {\bibinfo  {journal} {Phys. Rev. Lett.}\ }\textbf
  {\bibinfo {volume} {124}},\ \bibinfo {pages} {140502} (\bibinfo {year}
  {2020})}\BibitemShut {NoStop}%
\bibitem [{\citenamefont {Endo}\ \emph {et~al.}(2020)\citenamefont {Endo},
  \citenamefont {Sun}, \citenamefont {Li}, \citenamefont {Benjamin},\ and\
  \citenamefont {Yuan}}]{endo2020variational}%
  \BibitemOpen
  \bibfield  {author} {\bibinfo {author} {\bibfnamefont {S.}~\bibnamefont
  {Endo}}, \bibinfo {author} {\bibfnamefont {J.}~\bibnamefont {Sun}}, \bibinfo
  {author} {\bibfnamefont {Y.}~\bibnamefont {Li}}, \bibinfo {author}
  {\bibfnamefont {S.~C.}\ \bibnamefont {Benjamin}},\ and\ \bibinfo {author}
  {\bibfnamefont {X.}~\bibnamefont {Yuan}},\ }\href
  {https://doi.org/10.1103/PhysRevLett.125.010501} {\bibfield  {journal}
  {\bibinfo  {journal} {Phys. Rev. Lett.}\ }\textbf {\bibinfo {volume} {125}},\
  \bibinfo {pages} {010501} (\bibinfo {year} {2020})}\BibitemShut {NoStop}%
\bibitem [{\citenamefont {Schuld}\ \emph {et~al.}(2020)\citenamefont {Schuld},
  \citenamefont {Bocharov}, \citenamefont {Svore},\ and\ \citenamefont
  {Wiebe}}]{schuld2020circuit}%
  \BibitemOpen
  \bibfield  {author} {\bibinfo {author} {\bibfnamefont {M.}~\bibnamefont
  {Schuld}}, \bibinfo {author} {\bibfnamefont {A.}~\bibnamefont {Bocharov}},
  \bibinfo {author} {\bibfnamefont {K.~M.}\ \bibnamefont {Svore}},\ and\
  \bibinfo {author} {\bibfnamefont {N.}~\bibnamefont {Wiebe}},\ }\href
  {https://doi.org/10.1103/PhysRevA.101.032308} {\bibfield  {journal} {\bibinfo
   {journal} {Phys. Rev. A}\ }\textbf {\bibinfo {volume} {101}},\ \bibinfo
  {pages} {032308} (\bibinfo {year} {2020})}\BibitemShut {NoStop}%
\bibitem [{\citenamefont {Cerezo}\ \emph {et~al.}(2021)\citenamefont {Cerezo},
  \citenamefont {Arrasmith}, \citenamefont {Babbush}, \citenamefont {Benjamin},
  \citenamefont {Endo}, \citenamefont {Fujii}, \citenamefont {McClean},
  \citenamefont {Mitarai}, \citenamefont {Yuan}, \citenamefont {Cincio} \emph
  {et~al.}}]{cerezo2021variational}%
  \BibitemOpen
  \bibfield  {author} {\bibinfo {author} {\bibfnamefont {M.}~\bibnamefont
  {Cerezo}}, \bibinfo {author} {\bibfnamefont {A.}~\bibnamefont {Arrasmith}},
  \bibinfo {author} {\bibfnamefont {R.}~\bibnamefont {Babbush}}, \bibinfo
  {author} {\bibfnamefont {S.~C.}\ \bibnamefont {Benjamin}}, \bibinfo {author}
  {\bibfnamefont {S.}~\bibnamefont {Endo}}, \bibinfo {author} {\bibfnamefont
  {K.}~\bibnamefont {Fujii}}, \bibinfo {author} {\bibfnamefont {J.~R.}\
  \bibnamefont {McClean}}, \bibinfo {author} {\bibfnamefont {K.}~\bibnamefont
  {Mitarai}}, \bibinfo {author} {\bibfnamefont {X.}~\bibnamefont {Yuan}},
  \bibinfo {author} {\bibfnamefont {L.}~\bibnamefont {Cincio}}, \emph
  {et~al.},\ }\href
  {https://doi.org/https://doi.org/10.1038/s42254-021-00348-9} {\bibfield
  {journal} {\bibinfo  {journal} {Nat. Rev. Phys.}\ }\textbf {\bibinfo {volume}
  {3}},\ \bibinfo {pages} {625} (\bibinfo {year} {2021})}\BibitemShut {NoStop}%
\bibitem [{\citenamefont {Baldridge}\ and\ \citenamefont
  {Osborne}(2004)}]{baldridge2004active}%
  \BibitemOpen
  \bibfield  {author} {\bibinfo {author} {\bibfnamefont {J.}~\bibnamefont
  {Baldridge}}\ and\ \bibinfo {author} {\bibfnamefont {M.}~\bibnamefont
  {Osborne}},\ }in\ \href {https://doi.org/10.1.1.108.5493} {\emph {\bibinfo
  {booktitle} {Proceedings of the 2004 Conference on Empirical Methods in
  Natural Language Processing}}}\ (\bibinfo {year} {2004})\ pp.\ \bibinfo
  {pages} {9--16}\BibitemShut {NoStop}%
\bibitem [{\citenamefont {Settles}(2009)}]{settles2009active}%
  \BibitemOpen
  \bibfield  {author} {\bibinfo {author} {\bibfnamefont {B.}~\bibnamefont
  {Settles}}\ }\href
  {https://doi.org/http://digital.library.wisc.edu/1793/60660}
  {http://digital.library.wisc.edu/1793/60660} (\bibinfo {year}
  {2009})\BibitemShut {NoStop}%
\bibitem [{\citenamefont {Huang}\ \emph {et~al.}(2014)\citenamefont {Huang},
  \citenamefont {Jin},\ and\ \citenamefont {Zhou}}]{huang2014active}%
  \BibitemOpen
  \bibfield  {author} {\bibinfo {author} {\bibfnamefont {S.-J.}\ \bibnamefont
  {Huang}}, \bibinfo {author} {\bibfnamefont {R.}~\bibnamefont {Jin}},\ and\
  \bibinfo {author} {\bibfnamefont {Z.-H.}\ \bibnamefont {Zhou}},\ }\href
  {https://doi.org/10.1109/TPAMI.2014.2307881} {\bibfield  {journal} {\bibinfo
  {journal} {IEEE TPAMI}\ }\textbf {\bibinfo {volume} {36}},\ \bibinfo {pages}
  {1936} (\bibinfo {year} {2014})}\BibitemShut {NoStop}%
\bibitem [{\citenamefont {Ding}\ \emph {et~al.}(2020)\citenamefont {Ding},
  \citenamefont {Mart{\'\i}n-Guerrero}, \citenamefont {Sanz}, \citenamefont
  {Magdalena-Benedicto}, \citenamefont {Chen},\ and\ \citenamefont
  {Solano}}]{ding2020retrieving}%
  \BibitemOpen
  \bibfield  {author} {\bibinfo {author} {\bibfnamefont {Y.}~\bibnamefont
  {Ding}}, \bibinfo {author} {\bibfnamefont {J.~D.}\ \bibnamefont
  {Mart{\'\i}n-Guerrero}}, \bibinfo {author} {\bibfnamefont {M.}~\bibnamefont
  {Sanz}}, \bibinfo {author} {\bibfnamefont {R.}~\bibnamefont
  {Magdalena-Benedicto}}, \bibinfo {author} {\bibfnamefont {X.}~\bibnamefont
  {Chen}},\ and\ \bibinfo {author} {\bibfnamefont {E.}~\bibnamefont {Solano}},\
  }\href {https://doi.org/10.1103/PhysRevLett.124.140504} {\bibfield  {journal}
  {\bibinfo  {journal} {Phys. Rev. Lett.}\ }\textbf {\bibinfo {volume} {124}},\
  \bibinfo {pages} {140504} (\bibinfo {year} {2020})}\BibitemShut {NoStop}%
\bibitem [{\citenamefont {Dutt}\ \emph {et~al.}(2021)\citenamefont {Dutt},
  \citenamefont {Pednault}, \citenamefont {Wu}, \citenamefont {Sheldon},
  \citenamefont {Smolin}, \citenamefont {Bishop},\ and\ \citenamefont
  {Chuang}}]{dutt2021active}%
  \BibitemOpen
  \bibfield  {author} {\bibinfo {author} {\bibfnamefont {A.}~\bibnamefont
  {Dutt}}, \bibinfo {author} {\bibfnamefont {E.}~\bibnamefont {Pednault}},
  \bibinfo {author} {\bibfnamefont {C.~W.}\ \bibnamefont {Wu}}, \bibinfo
  {author} {\bibfnamefont {S.}~\bibnamefont {Sheldon}}, \bibinfo {author}
  {\bibfnamefont {J.}~\bibnamefont {Smolin}}, \bibinfo {author} {\bibfnamefont
  {L.}~\bibnamefont {Bishop}},\ and\ \bibinfo {author} {\bibfnamefont {I.~L.}\
  \bibnamefont {Chuang}},\ }\bibfield  {journal} {\bibinfo  {journal}
  {arXiv:2112.14553}\ }\href
  {https://doi.org/https://doi.org/10.48550/arXiv.2112.14553}
  {https://doi.org/10.48550/arXiv.2112.14553} (\bibinfo {year}
  {2021})\BibitemShut {NoStop}%
\bibitem [{\citenamefont {Ding}\ \emph {et~al.}(2022)\citenamefont {Ding},
  \citenamefont {Mart\'{\i}n-Guerrero}, \citenamefont {Song}, \citenamefont
  {Magdalena-Benedicto},\ and\ \citenamefont {Chen}}]{ding2022active}%
  \BibitemOpen
  \bibfield  {author} {\bibinfo {author} {\bibfnamefont {Y.}~\bibnamefont
  {Ding}}, \bibinfo {author} {\bibfnamefont {J.~D.}\ \bibnamefont
  {Mart\'{\i}n-Guerrero}}, \bibinfo {author} {\bibfnamefont {Y.}~\bibnamefont
  {Song}}, \bibinfo {author} {\bibfnamefont {R.}~\bibnamefont
  {Magdalena-Benedicto}},\ and\ \bibinfo {author} {\bibfnamefont
  {X.}~\bibnamefont {Chen}},\ }\href
  {https://doi.org/10.1103/PhysRevResearch.4.013213} {\bibfield  {journal}
  {\bibinfo  {journal} {Phys. Rev. Research}\ }\textbf {\bibinfo {volume}
  {4}},\ \bibinfo {pages} {013213} (\bibinfo {year} {2022})}\BibitemShut
  {NoStop}%
\bibitem [{\citenamefont {Melnikov}\ \emph {et~al.}(2018)\citenamefont
  {Melnikov}, \citenamefont {Poulsen~Nautrup}, \citenamefont {Krenn},
  \citenamefont {Dunjko}, \citenamefont {Tiersch}, \citenamefont {Zeilinger},\
  and\ \citenamefont {Briegel}}]{melnikov2018active}%
  \BibitemOpen
  \bibfield  {author} {\bibinfo {author} {\bibfnamefont {A.~A.}\ \bibnamefont
  {Melnikov}}, \bibinfo {author} {\bibfnamefont {H.}~\bibnamefont
  {Poulsen~Nautrup}}, \bibinfo {author} {\bibfnamefont {M.}~\bibnamefont
  {Krenn}}, \bibinfo {author} {\bibfnamefont {V.}~\bibnamefont {Dunjko}},
  \bibinfo {author} {\bibfnamefont {M.}~\bibnamefont {Tiersch}}, \bibinfo
  {author} {\bibfnamefont {A.}~\bibnamefont {Zeilinger}},\ and\ \bibinfo
  {author} {\bibfnamefont {H.~J.}\ \bibnamefont {Briegel}},\ }\href
  {https://doi.org/10.1073/pnas.1714936115} {\bibfield  {journal} {\bibinfo
  {journal} {Proc. Natl. Acad. Sci. U. S. A.}\ }\textbf {\bibinfo {volume}
  {115}},\ \bibinfo {pages} {1221} (\bibinfo {year} {2018})}\BibitemShut
  {NoStop}%
\bibitem [{\citenamefont {Settles}\ \emph {et~al.}(2008)\citenamefont
  {Settles}, \citenamefont {Craven},\ and\ \citenamefont
  {Friedland}}]{settles2008active}%
  \BibitemOpen
  \bibfield  {author} {\bibinfo {author} {\bibfnamefont {B.}~\bibnamefont
  {Settles}}, \bibinfo {author} {\bibfnamefont {M.}~\bibnamefont {Craven}},\
  and\ \bibinfo {author} {\bibfnamefont {L.}~\bibnamefont {Friedland}},\ }in\
  \href {https://doi.org/10.1.1.329.8090} {\emph {\bibinfo {booktitle}
  {Proceedings of the NIPS workshop on cost-sensitive learning}}},\
  Vol.~\bibinfo {volume} {1}\ (\bibinfo {organization} {Vancouver, CA:},\
  \bibinfo {year} {2008})\BibitemShut {NoStop}%
\bibitem [{\citenamefont {Lewis}\ and\ \citenamefont
  {Gale}(1994)}]{lewis1994sequential}%
  \BibitemOpen
  \bibfield  {author} {\bibinfo {author} {\bibfnamefont {D.~D.}\ \bibnamefont
  {Lewis}}\ and\ \bibinfo {author} {\bibfnamefont {W.~A.}\ \bibnamefont
  {Gale}},\ }in\ \href
  {https://doi.org/https://doi.org/10.1007/978-1-4471-2099-5_1} {\emph
  {\bibinfo {booktitle} {SIGIR’94}}}\ (\bibinfo {organization} {Springer},\
  \bibinfo {year} {1994})\ pp.\ \bibinfo {pages} {3--12}\BibitemShut {NoStop}%
\bibitem [{\citenamefont {Scheffer}\ \emph {et~al.}(2001)\citenamefont
  {Scheffer}, \citenamefont {Decomain},\ and\ \citenamefont
  {Wrobel}}]{scheffer2001active}%
  \BibitemOpen
  \bibfield  {author} {\bibinfo {author} {\bibfnamefont {T.}~\bibnamefont
  {Scheffer}}, \bibinfo {author} {\bibfnamefont {C.}~\bibnamefont {Decomain}},\
  and\ \bibinfo {author} {\bibfnamefont {S.}~\bibnamefont {Wrobel}},\ }in\
  \href {https://doi.org/https://doi.org/10.1007/3-540-44816-0_31} {\emph
  {\bibinfo {booktitle} {Proceedings of the 4th International Conference on
  Advances in Intelligent Data Analysis}}},\ \bibinfo {series and number} {IDA
  '01}\ (\bibinfo  {publisher} {Springer-Verlag},\ \bibinfo {address} {Berlin,
  Heidelberg},\ \bibinfo {year} {2001})\ p.\ \bibinfo {pages}
  {309–318}\BibitemShut {NoStop}%
\bibitem [{\citenamefont {Seung}\ \emph {et~al.}(1992)\citenamefont {Seung},
  \citenamefont {Opper},\ and\ \citenamefont {Sompolinsky}}]{seung1992query}%
  \BibitemOpen
  \bibfield  {author} {\bibinfo {author} {\bibfnamefont {H.~S.}\ \bibnamefont
  {Seung}}, \bibinfo {author} {\bibfnamefont {M.}~\bibnamefont {Opper}},\ and\
  \bibinfo {author} {\bibfnamefont {H.}~\bibnamefont {Sompolinsky}},\ }in\
  \href {https://doi.org/https://doi.org/10.1145/130385.130417} {\emph
  {\bibinfo {booktitle} {Proceedings of the fifth annual workshop on
  Computational learning theory}}}\ (\bibinfo {year} {1992})\ pp.\ \bibinfo
  {pages} {287--294}\BibitemShut {NoStop}%
\bibitem [{\citenamefont {Settles}\ \emph {et~al.}(2007)\citenamefont
  {Settles}, \citenamefont {Craven},\ and\ \citenamefont
  {Ray}}]{settles2007advances}%
  \BibitemOpen
  \bibfield  {author} {\bibinfo {author} {\bibfnamefont {B.}~\bibnamefont
  {Settles}}, \bibinfo {author} {\bibfnamefont {M.}~\bibnamefont {Craven}},\
  and\ \bibinfo {author} {\bibfnamefont {S.}~\bibnamefont {Ray}},\ }in\ \href
  {https://doi.org/https://dl.acm.org/doi/abs/10.5555/2981562.2981724} {\emph
  {\bibinfo {booktitle} {Advances in Neural Information Processing Systems}}},\
  Vol.~\bibinfo {volume} {20},\ \bibinfo {editor} {edited by\ \bibinfo {editor}
  {\bibfnamefont {J.}~\bibnamefont {Platt}}, \bibinfo {editor} {\bibfnamefont
  {D.}~\bibnamefont {Koller}}, \bibinfo {editor} {\bibfnamefont
  {Y.}~\bibnamefont {Singer}},\ and\ \bibinfo {editor} {\bibfnamefont
  {S.}~\bibnamefont {Roweis}}}\ (\bibinfo  {publisher} {Curran Associates,
  Inc.},\ \bibinfo {year} {2007})\BibitemShut {NoStop}%
\bibitem [{\citenamefont {Roy}\ and\ \citenamefont
  {McCallum}(2001)}]{roy2001toward}%
  \BibitemOpen
  \bibfield  {author} {\bibinfo {author} {\bibfnamefont {N.}~\bibnamefont
  {Roy}}\ and\ \bibinfo {author} {\bibfnamefont {A.}~\bibnamefont {McCallum}},\
  }\bibfield  {journal} {\bibinfo  {journal} {Int. Conf. on Mach. Learn.}\
  }\href {https://doi.org/https://dl.acm.org/doi/10.5555/645530.655646}
  {https://dl.acm.org/doi/10.5555/645530.655646} (\bibinfo {year}
  {2001})\BibitemShut {NoStop}%
\bibitem [{\citenamefont {Fuchs}(1996)}]{fuchs1996distinguishability}%
  \BibitemOpen
  \bibfield  {author} {\bibinfo {author} {\bibfnamefont {C.~A.}\ \bibnamefont
  {Fuchs}},\ }\bibfield  {journal} {\bibinfo  {journal} {University of New
  Mexico, Albuquerque}\ }\href
  {https://doi.org/https://doi.org/10.48550/arXiv.quant-ph/9601020}
  {https://doi.org/10.48550/arXiv.quant-ph/9601020} (\bibinfo {year}
  {1996})\BibitemShut {NoStop}%
\bibitem [{\citenamefont {Schuld}\ \emph {et~al.}(2019)\citenamefont {Schuld},
  \citenamefont {Bergholm}, \citenamefont {Gogolin}, \citenamefont {Izaac},\
  and\ \citenamefont {Killoran}}]{schuld2019evaluating}%
  \BibitemOpen
  \bibfield  {author} {\bibinfo {author} {\bibfnamefont {M.}~\bibnamefont
  {Schuld}}, \bibinfo {author} {\bibfnamefont {V.}~\bibnamefont {Bergholm}},
  \bibinfo {author} {\bibfnamefont {C.}~\bibnamefont {Gogolin}}, \bibinfo
  {author} {\bibfnamefont {J.}~\bibnamefont {Izaac}},\ and\ \bibinfo {author}
  {\bibfnamefont {N.}~\bibnamefont {Killoran}},\ }\href
  {https://dx.doi.org/10.1103/PhysRevA.99.032331} {\bibfield  {journal}
  {\bibinfo  {journal} {Phys. Rev. A}\ }\textbf {\bibinfo {volume} {99}}
  (\bibinfo {year} {2019})}\BibitemShut {NoStop}%
\bibitem [{\citenamefont {Wang}\ \emph {et~al.}(2016)\citenamefont {Wang},
  \citenamefont {Chen}, \citenamefont {Li}, \citenamefont {Huang},
  \citenamefont {Liu}, \citenamefont {Chen}, \citenamefont {Luo}, \citenamefont
  {Su}, \citenamefont {Wu}, \citenamefont {Li} \emph
  {et~al.}}]{wang2016experimental}%
  \BibitemOpen
  \bibfield  {author} {\bibinfo {author} {\bibfnamefont {X.-L.}\ \bibnamefont
  {Wang}}, \bibinfo {author} {\bibfnamefont {L.-K.}\ \bibnamefont {Chen}},
  \bibinfo {author} {\bibfnamefont {W.}~\bibnamefont {Li}}, \bibinfo {author}
  {\bibfnamefont {H.-L.}\ \bibnamefont {Huang}}, \bibinfo {author}
  {\bibfnamefont {C.}~\bibnamefont {Liu}}, \bibinfo {author} {\bibfnamefont
  {C.}~\bibnamefont {Chen}}, \bibinfo {author} {\bibfnamefont {Y.-H.}\
  \bibnamefont {Luo}}, \bibinfo {author} {\bibfnamefont {Z.-E.}\ \bibnamefont
  {Su}}, \bibinfo {author} {\bibfnamefont {D.}~\bibnamefont {Wu}}, \bibinfo
  {author} {\bibfnamefont {Z.-D.}\ \bibnamefont {Li}}, \emph {et~al.},\ }\href
  {https://doi.org/10.1103/PhysRevLett.117.210502} {\bibfield  {journal}
  {\bibinfo  {journal} {Phys. Rev. Lett.}\ }\textbf {\bibinfo {volume} {117}},\
  \bibinfo {pages} {210502} (\bibinfo {year} {2016})}\BibitemShut {NoStop}%
\bibitem [{\citenamefont {Kingma}\ and\ \citenamefont
  {Ba}(2014)}]{kingma2015Adam}%
  \BibitemOpen
  \bibfield  {author} {\bibinfo {author} {\bibfnamefont {D.~P.}\ \bibnamefont
  {Kingma}}\ and\ \bibinfo {author} {\bibfnamefont {J.}~\bibnamefont {Ba}},\
  }\bibfield  {journal} {\bibinfo  {journal} {arXiv:1412.6980}\ }\href
  {https://doi.org/https://doi.org/10.48550/arXiv.1412.6980}
  {https://doi.org/10.48550/arXiv.1412.6980} (\bibinfo {year}
  {2014})\BibitemShut {NoStop}%
\bibitem [{\citenamefont {Drexl}\ and\ \citenamefont
  {Schneider}(2015)}]{drexl2015survey}%
  \BibitemOpen
  \bibfield  {author} {\bibinfo {author} {\bibfnamefont {M.}~\bibnamefont
  {Drexl}}\ and\ \bibinfo {author} {\bibfnamefont {M.}~\bibnamefont
  {Schneider}},\ }\href
  {https://doi.org/https://doi.org/10.1016/j.ejor.2014.08.030} {\bibfield
  {journal} {\bibinfo  {journal} {Eur. J. Oper. Res.}\ }\textbf {\bibinfo
  {volume} {241}},\ \bibinfo {pages} {283} (\bibinfo {year}
  {2015})}\BibitemShut {NoStop}%
\bibitem [{\citenamefont {Cortes~Corinna}(1995)}]{cortes1995support}%
  \BibitemOpen
  \bibfield  {author} {\bibinfo {author} {\bibfnamefont {V.~V.}\ \bibnamefont
  {Cortes~Corinna}},\ }\href {https://doi.org/10.1023/A:1022627411411}
  {\bibfield  {journal} {\bibinfo  {journal} {Mach. Learn.}\ }\textbf {\bibinfo
  {volume} {20}},\ \bibinfo {pages} {273} (\bibinfo {year} {1995})}\BibitemShut
  {NoStop}%
\bibitem [{\citenamefont {Keerthi}\ and\ \citenamefont
  {Lin}(2003)}]{Keerthi2003asymptotic}%
  \BibitemOpen
  \bibfield  {author} {\bibinfo {author} {\bibfnamefont {S.~S.}\ \bibnamefont
  {Keerthi}}\ and\ \bibinfo {author} {\bibfnamefont {C.-J.}\ \bibnamefont
  {Lin}},\ }\href {https://doi.org/10.1162/089976603321891855} {\bibfield
  {journal} {\bibinfo  {journal} {Neural Comput.}\ }\textbf {\bibinfo {volume}
  {15}},\ \bibinfo {pages} {1667} (\bibinfo {year} {2003})}\BibitemShut
  {NoStop}%
\bibitem [{\citenamefont {Cover}\ and\ \citenamefont
  {Hart}(1967)}]{cover1967nearest}%
  \BibitemOpen
  \bibfield  {author} {\bibinfo {author} {\bibfnamefont {T.}~\bibnamefont
  {Cover}}\ and\ \bibinfo {author} {\bibfnamefont {P.}~\bibnamefont {Hart}},\
  }\href {https://doi.org/10.1109/TIT.1967.1053964} {\bibfield  {journal}
  {\bibinfo  {journal} {IEEE TIT}\ }\textbf {\bibinfo {volume} {13}},\ \bibinfo
  {pages} {21} (\bibinfo {year} {1967})}\BibitemShut {NoStop}%
\bibitem [{\citenamefont {Bailey}\ and\ \citenamefont
  {AK}(1978)}]{bailey1978anote}%
  \BibitemOpen
  \bibfield  {author} {\bibinfo {author} {\bibfnamefont {T.}~\bibnamefont
  {Bailey}}\ and\ \bibinfo {author} {\bibfnamefont {J.}~\bibnamefont {AK}},\
  }\href {https://doi.org/10.1109/TSMC.1978.4309958} {\bibfield  {journal}
  {\bibinfo  {journal} {IEEE Trans. Syst. Man Cybern.}\ }\textbf {\bibinfo
  {volume} {8}},\ \bibinfo {pages} {311} (\bibinfo {year} {1978})}\BibitemShut
  {NoStop}%
\bibitem [{\citenamefont {Fisher}(1936)}]{fisher1936theuse}%
  \BibitemOpen
  \bibfield  {author} {\bibinfo {author} {\bibfnamefont {R.~A.}\ \bibnamefont
  {Fisher}},\ }\href
  {https://doi.org/https://doi.org/10.1111/j.1469-1809.1936.tb02137.x}
  {\bibfield  {journal} {\bibinfo  {journal} {Ann. Eugen.}\ }\textbf {\bibinfo
  {volume} {7}},\ \bibinfo {pages} {179} (\bibinfo {year} {1936})}\BibitemShut
  {NoStop}%
\bibitem [{\citenamefont {Huberty}(1975)}]{carl1975discriminant}%
  \BibitemOpen
  \bibfield  {author} {\bibinfo {author} {\bibfnamefont {C.~J.}\ \bibnamefont
  {Huberty}},\ }\href {https://doi.org/10.3102/00346543045004543} {\bibfield
  {journal} {\bibinfo  {journal} {Educ. Res. Rev.}\ }\textbf {\bibinfo {volume}
  {45}},\ \bibinfo {pages} {543} (\bibinfo {year} {1975})}\BibitemShut
  {NoStop}%
\bibitem [{\citenamefont {Quinlan}(1986)}]{quinlan1986induction}%
  \BibitemOpen
  \bibfield  {author} {\bibinfo {author} {\bibfnamefont {J.~R.}\ \bibnamefont
  {Quinlan}},\ }\href {https://doi.org/10.1007/BF00116251} {\bibfield
  {journal} {\bibinfo  {journal} {Mach. Learn.}\ }\textbf {\bibinfo {volume}
  {1}},\ \bibinfo {pages} {81} (\bibinfo {year} {1986})}\BibitemShut {NoStop}%
\bibitem [{\citenamefont {Safavian}\ and\ \citenamefont
  {Landgrebe}(1991)}]{safavian1991asurvey}%
  \BibitemOpen
  \bibfield  {author} {\bibinfo {author} {\bibfnamefont {S.}~\bibnamefont
  {Safavian}}\ and\ \bibinfo {author} {\bibfnamefont {D.}~\bibnamefont
  {Landgrebe}},\ }\href {https://doi.org/10.1109/21.97458} {\bibfield
  {journal} {\bibinfo  {journal} {IEEE Trans. Syst. Man Cybern.}\ }\textbf
  {\bibinfo {volume} {21}},\ \bibinfo {pages} {660} (\bibinfo {year}
  {1991})}\BibitemShut {NoStop}%
\bibitem [{\citenamefont {Aaronson}(2005)}]{aaronson2005quantum}%
  \BibitemOpen
  \bibfield  {author} {\bibinfo {author} {\bibfnamefont {S.}~\bibnamefont
  {Aaronson}},\ }\href {https://doi.org/https://doi.org/10.1098/rspa.2005.1546}
  {\bibfield  {journal} {\bibinfo  {journal} {Proc. R. Soc. A.}\ }\textbf
  {\bibinfo {volume} {461}},\ \bibinfo {pages} {3473} (\bibinfo {year}
  {2005})}\BibitemShut {NoStop}%
\bibitem [{\citenamefont {Holmes}\ \emph {et~al.}(2021)\citenamefont {Holmes},
  \citenamefont {Coble}, \citenamefont {Sornborger},\ and\ \citenamefont
  {Suba{\c{s}}{\i}}}]{holmes2021nonlinear}%
  \BibitemOpen
  \bibfield  {author} {\bibinfo {author} {\bibfnamefont {Z.}~\bibnamefont
  {Holmes}}, \bibinfo {author} {\bibfnamefont {N.}~\bibnamefont {Coble}},
  \bibinfo {author} {\bibfnamefont {A.~T.}\ \bibnamefont {Sornborger}},\ and\
  \bibinfo {author} {\bibfnamefont {Y.}~\bibnamefont {Suba{\c{s}}{\i}}},\
  }\bibfield  {journal} {\bibinfo  {journal} {arXiv:2112.12307}\ }\href
  {https://doi.org/https://doi.org/10.48550/arXiv.2112.12307}
  {https://doi.org/10.48550/arXiv.2112.12307} (\bibinfo {year}
  {2021})\BibitemShut {NoStop}%
\bibitem [{\citenamefont {Zhong}\ \emph {et~al.}(2020)\citenamefont {Zhong},
  \citenamefont {Wang}, \citenamefont {Deng}, \citenamefont {Chen},
  \citenamefont {Peng}, \citenamefont {Luo}, \citenamefont {Qin}, \citenamefont
  {Wu}, \citenamefont {Ding}, \citenamefont {Hu} \emph
  {et~al.}}]{zhong2020quantum}%
  \BibitemOpen
  \bibfield  {author} {\bibinfo {author} {\bibfnamefont {H.-S.}\ \bibnamefont
  {Zhong}}, \bibinfo {author} {\bibfnamefont {H.}~\bibnamefont {Wang}},
  \bibinfo {author} {\bibfnamefont {Y.-H.}\ \bibnamefont {Deng}}, \bibinfo
  {author} {\bibfnamefont {M.-C.}\ \bibnamefont {Chen}}, \bibinfo {author}
  {\bibfnamefont {L.-C.}\ \bibnamefont {Peng}}, \bibinfo {author}
  {\bibfnamefont {Y.-H.}\ \bibnamefont {Luo}}, \bibinfo {author} {\bibfnamefont
  {J.}~\bibnamefont {Qin}}, \bibinfo {author} {\bibfnamefont {D.}~\bibnamefont
  {Wu}}, \bibinfo {author} {\bibfnamefont {X.}~\bibnamefont {Ding}}, \bibinfo
  {author} {\bibfnamefont {Y.}~\bibnamefont {Hu}}, \emph {et~al.},\ }\href
  {https://doi.org/10.1126/science.abe8770} {\bibfield  {journal} {\bibinfo
  {journal} {Science}\ }\textbf {\bibinfo {volume} {370}},\ \bibinfo {pages}
  {1460} (\bibinfo {year} {2020})}\BibitemShut {NoStop}%
\bibitem [{\citenamefont {Zhong}\ \emph {et~al.}(2021)\citenamefont {Zhong},
  \citenamefont {Deng}, \citenamefont {Qin}, \citenamefont {Wang},
  \citenamefont {Chen}, \citenamefont {Peng}, \citenamefont {Luo},
  \citenamefont {Wu}, \citenamefont {Gong}, \citenamefont {Su} \emph
  {et~al.}}]{zhong2021phase}%
  \BibitemOpen
  \bibfield  {author} {\bibinfo {author} {\bibfnamefont {H.-S.}\ \bibnamefont
  {Zhong}}, \bibinfo {author} {\bibfnamefont {Y.-H.}\ \bibnamefont {Deng}},
  \bibinfo {author} {\bibfnamefont {J.}~\bibnamefont {Qin}}, \bibinfo {author}
  {\bibfnamefont {H.}~\bibnamefont {Wang}}, \bibinfo {author} {\bibfnamefont
  {M.-C.}\ \bibnamefont {Chen}}, \bibinfo {author} {\bibfnamefont {L.-C.}\
  \bibnamefont {Peng}}, \bibinfo {author} {\bibfnamefont {Y.-H.}\ \bibnamefont
  {Luo}}, \bibinfo {author} {\bibfnamefont {D.}~\bibnamefont {Wu}}, \bibinfo
  {author} {\bibfnamefont {S.-Q.}\ \bibnamefont {Gong}}, \bibinfo {author}
  {\bibfnamefont {H.}~\bibnamefont {Su}}, \emph {et~al.},\ }\href
  {https://doi.org/10.1103/PhysRevLett.127.180502} {\bibfield  {journal}
  {\bibinfo  {journal} {Phys. Rev. Lett.}\ }\textbf {\bibinfo {volume} {127}},\
  \bibinfo {pages} {180502} (\bibinfo {year} {2021})}\BibitemShut {NoStop}%
\end{thebibliography}%


\begin{thebibliography}{8}%
\makeatletter
\providecommand \@ifxundefined [1]{%
 \@ifx{#1\undefined}
}%
\providecommand \@ifnum [1]{%
 \ifnum #1\expandafter \@firstoftwo
 \else \expandafter \@secondoftwo
 \fi
}%
\providecommand \@ifx [1]{%
 \ifx #1\expandafter \@firstoftwo
 \else \expandafter \@secondoftwo
 \fi
}%
\providecommand \natexlab [1]{#1}%
\providecommand \enquote  [1]{``#1''}%
\providecommand \bibnamefont  [1]{#1}%
\providecommand \bibfnamefont [1]{#1}%
\providecommand \citenamefont [1]{#1}%
\providecommand \href@noop [0]{\@secondoftwo}%
\providecommand \href [0]{\begingroup \@sanitize@url \@href}%
\providecommand \@href[1]{\@@startlink{#1}\@@href}%
\providecommand \@@href[1]{\endgroup#1\@@endlink}%
\providecommand \@sanitize@url [0]{\catcode `\\12\catcode `\$12\catcode
  `\&12\catcode `\#12\catcode `\^12\catcode `\_12\catcode `\%12\relax}%
\providecommand \@@startlink[1]{}%
\providecommand \@@endlink[0]{}%
\providecommand \url  [0]{\begingroup\@sanitize@url \@url }%
\providecommand \@url [1]{\endgroup\@href {#1}{\urlprefix }}%
\providecommand \urlprefix  [0]{URL }%
\providecommand \Eprint [0]{\href }%
\providecommand \doibase [0]{https://doi.org/}%
\providecommand \selectlanguage [0]{\@gobble}%
\providecommand \bibinfo  [0]{\@secondoftwo}%
\providecommand \bibfield  [0]{\@secondoftwo}%
\providecommand \translation [1]{[#1]}%
\providecommand \BibitemOpen [0]{}%
\providecommand \bibitemStop [0]{}%
\providecommand \bibitemNoStop [0]{.\EOS\space}%
\providecommand \EOS [0]{\spacefactor3000\relax}%
\providecommand \BibitemShut  [1]{\csname bibitem#1\endcsname}%
\let\auto@bib@innerbib\@empty
\bibitem [{\citenamefont {Cortes~Corinna}(1995)}]{cortes1995support}%
  \BibitemOpen
  \bibfield  {author} {\bibinfo {author} {\bibfnamefont {V.~V.}\ \bibnamefont
  {Cortes~Corinna}},\ }\href {https://doi.org/10.1023/A:1022627411411}
  {\bibfield  {journal} {\bibinfo  {journal} {Mach. Learn.}\ }\textbf {\bibinfo
  {volume} {20}},\ \bibinfo {pages} {273} (\bibinfo {year} {1995})}\BibitemShut
  {NoStop}%
\bibitem [{\citenamefont {Keerthi}\ and\ \citenamefont
  {Lin}(2003)}]{Keerthi2003asymptotic}%
  \BibitemOpen
  \bibfield  {author} {\bibinfo {author} {\bibfnamefont {S.~S.}\ \bibnamefont
  {Keerthi}}\ and\ \bibinfo {author} {\bibfnamefont {C.-J.}\ \bibnamefont
  {Lin}},\ }\href {https://doi.org/10.1162/089976603321891855} {\bibfield
  {journal} {\bibinfo  {journal} {Neural Comput.}\ }\textbf {\bibinfo {volume}
  {15}},\ \bibinfo {pages} {1667} (\bibinfo {year} {2003})}\BibitemShut
  {NoStop}%
\bibitem [{\citenamefont {Cover}\ and\ \citenamefont
  {Hart}(1967)}]{cover1967nearest}%
  \BibitemOpen
  \bibfield  {author} {\bibinfo {author} {\bibfnamefont {T.}~\bibnamefont
  {Cover}}\ and\ \bibinfo {author} {\bibfnamefont {P.}~\bibnamefont {Hart}},\
  }\href {https://doi.org/10.1109/TIT.1967.1053964} {\bibfield  {journal}
  {\bibinfo  {journal} {IEEE TIT}\ }\textbf {\bibinfo {volume} {13}},\ \bibinfo
  {pages} {21} (\bibinfo {year} {1967})}\BibitemShut {NoStop}%
\bibitem [{\citenamefont {Bailey}\ and\ \citenamefont
  {AK}(1978)}]{bailey1978anote}%
  \BibitemOpen
  \bibfield  {author} {\bibinfo {author} {\bibfnamefont {T.}~\bibnamefont
  {Bailey}}\ and\ \bibinfo {author} {\bibfnamefont {J.}~\bibnamefont {AK}},\
  }\href {https://doi.org/10.1109/TSMC.1978.4309958} {\bibfield  {journal}
  {\bibinfo  {journal} {IEEE Trans. Syst. Man Cybern.}\ }\textbf {\bibinfo
  {volume} {8}},\ \bibinfo {pages} {311} (\bibinfo {year} {1978})}\BibitemShut
  {NoStop}%
\bibitem [{\citenamefont {Fisher}(1936)}]{fisher1936theuse}%
  \BibitemOpen
  \bibfield  {author} {\bibinfo {author} {\bibfnamefont {R.~A.}\ \bibnamefont
  {Fisher}},\ }\href
  {https://doi.org/https://doi.org/10.1111/j.1469-1809.1936.tb02137.x}
  {\bibfield  {journal} {\bibinfo  {journal} {Ann. Eugen.}\ }\textbf {\bibinfo
  {volume} {7}},\ \bibinfo {pages} {179} (\bibinfo {year} {1936})}\BibitemShut
  {NoStop}%
\bibitem [{\citenamefont {Huberty}(1975)}]{carl1975discriminant}%
  \BibitemOpen
  \bibfield  {author} {\bibinfo {author} {\bibfnamefont {C.~J.}\ \bibnamefont
  {Huberty}},\ }\href {https://doi.org/10.3102/00346543045004543} {\bibfield
  {journal} {\bibinfo  {journal} {Educ. Res. Rev.}\ }\textbf {\bibinfo {volume}
  {45}},\ \bibinfo {pages} {543} (\bibinfo {year} {1975})}\BibitemShut
  {NoStop}%
\bibitem [{\citenamefont {Quinlan}(1986)}]{quinlan1986induction}%
  \BibitemOpen
  \bibfield  {author} {\bibinfo {author} {\bibfnamefont {J.~R.}\ \bibnamefont
  {Quinlan}},\ }\href {https://doi.org/10.1007/BF00116251} {\bibfield
  {journal} {\bibinfo  {journal} {Mach. Learn.}\ }\textbf {\bibinfo {volume}
  {1}},\ \bibinfo {pages} {81} (\bibinfo {year} {1986})}\BibitemShut {NoStop}%
\bibitem [{\citenamefont {Safavian}\ and\ \citenamefont
  {Landgrebe}(1991)}]{safavian1991asurvey}%
  \BibitemOpen
  \bibfield  {author} {\bibinfo {author} {\bibfnamefont {S.}~\bibnamefont
  {Safavian}}\ and\ \bibinfo {author} {\bibfnamefont {D.}~\bibnamefont
  {Landgrebe}},\ }\href {https://doi.org/10.1109/21.97458} {\bibfield
  {journal} {\bibinfo  {journal} {IEEE Trans. Syst. Man Cybern.}\ }\textbf
  {\bibinfo {volume} {21}},\ \bibinfo {pages} {660} (\bibinfo {year}
  {1991})}\BibitemShut {NoStop}%
\end{thebibliography}%
\bibliographystyle{apsrev4-2}

\end{document}


\title{Supplementary information to "Active Learning on a Programmable Photonic Quantum Processor" }
\maketitle

\section{members of committee in QBC selection strategy}
The Query-by-Committee (QBC) is one of the selection strategies to evaluate the importance of data in our active learning experiments. The QBC committee consists of four members. They are SVC with the RBF kernel, $k$-nearest neighbor with $k=3$, Linear Discriminant Analysis algorithm (LDA), and decision tree with max depth 7.

SVC~\cite{cortes1995support, Keerthi2003asymptotic} is a supervised binary classification model. It learns a boundary that leads to the max margin from both classes of data. Rather than fitting all the training data, it focuses on the points close to the boundary, which named support vectors. Combined with the kernel function, SVC can express a very complex classification boundary to achieve better classification performance. The kernel function is a special nonlinear function which maps the low-dimensional feature space to a higher one. So the original linearly inseparable problem in the low-dimensional space transforms to that linearly separable in the high-dimensional space. The RBF kernel is one of typical kernel functions which we choose to enhance our SVC model.

KNN~\cite{cover1967nearest, bailey1978anote} is one of the most classic and simplest supervised machine learning models, which contains no explicit learning or training procedures. KNN concentrates on the $k$ nearset data points to the test data. If most nerighbor data points belong to a certain class, the test data is also divided into this class. The hyperparameters in the KNN algorithm are the distance metric and the $k$ value. We use the Euclidean distance metric and set the $k$ value to 3.

LDA~\cite{fisher1936theuse, carl1975discriminant}, also called Fisher’s linear discriminant, is a fundamental supervised dimensionality reduction method. The key idea of LDA is to maximize the between-class variance and minimize the within-class variance. Take binary classification as an example, LDA fisrt projects the data points in the feature space onto a line. Then it learns to optimize the line to make the projections of data points in the same class on the line as close as possible. Meanwhile, for projections of data points in different classes, the distances are maximized. To predict the test data, we calculate the projection of the test data point on the optimized line. Then the estimated label is judged according to the position of the projection.

Decision tree algorithm~\cite{quinlan1986induction, safavian1991asurvey} is a common supervised learning algorithm. It depends on a directed decision tree to make decisions on multi-classification task. During the training process, the decision tree is built according to the principle of minimizing the loss function. A decision tree consists of nodes and directed edges. There are two types of nodes: internal node (records a feature of dataset) and leaf node (denotes a class). Each directed edge corresponds to the certain value of a feature. When predicting a test data, which branch to enter is determined according to the judgment result of the internal node. Until the test data reaches the leaf node, we obtain the final classification result.  

\section{The classification power of VQC and NEVQC}
In this section, we provide rigorous derivation of the theoretical upper bound on the classification accuracy of the two classifiers.
\begin{figure*}
\begin{center}
\includegraphics[width=0.7\linewidth]{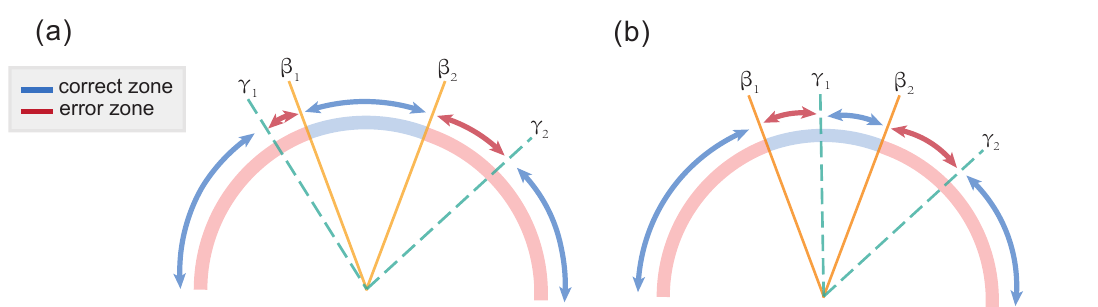}
\end{center}
\caption{\textbf{The illustration for two classification situations on an arc.} The green dashed line with slope angle $\gamma_1$ or $\gamma_2$ represents the classification line at current model parameters. The orange solid line with slope angle $\beta_1$ or $\beta_2$ refers to the actual dividing margin of a certain pattern. There are two general classification situations appearing in the training process. (a) Two classification lines cross through data points in the same class. In this situation, data items of at least one class are exactly classified. (b) Two classification lines cross through data points in different classes. Moreover, the blue arrows indicate the correctly classified zone and the red arrows point to the error zone.}
\label{fig-ability}
\end{figure*}

\subsection{Variational Quantum Classifier}
The implementation of variational quantum classifier(VQC) contains only one parameterized single-qubit gate formed by the HWP. By evaluating the sign of the measured $Z$ expectation, VQC actually determines the predicted label through a comparison bewteen probability $\text{P}_0$ and $\text{P}_1$, where $\text{P}_0$ and $\text{P}_1$ represent the probability of $|0\rangle$ and $|1\rangle$ events in the detectors. To obtain the analytical form of the probability, we consider the transformation within the classifier.
\begin{equation}
\begin{aligned}
|\psi\rangle_{\text{en}}^{\text{VQC}} &= x_i|0\rangle + y_i|1\rangle \\
&\xrightarrow[]{\text{HWP}(\theta)} ({x_i}\cos2\theta + {y_i}\sin2\theta) |0\rangle + ({x_i}\sin2\theta - {y_i}\cos2\theta) |1\rangle,
\end{aligned}
\end{equation}
where $|\psi\rangle_{\text{en}}^{\text{VQC}}$ refers to the input state of VQC and ($x_i$, $y_i$) is the encoding data. 

For the case of $\langle Z\rangle  > 0$, we have

\begin{equation}
\begin{aligned}
({x_i}\cos2\theta + {y_i}\sin2\theta)^2 
> ({x_i}\sin2\theta - {y_i}\cos2\theta)^2 
\end{aligned}.
\end{equation}
With a simple deformation, the inequality is re-expressed as
\begin{equation}
\begin{aligned}
({x_i}\sin(2\theta + \frac{\pi}{4}) + {y_i}\sin(2\theta - \frac{\pi}{4}))
\cdot(-{x_i}\sin(2\theta - \frac{\pi}{4}) + {y_i}\sin(2\theta + \frac{\pi}{4})) > 0
\end{aligned}.
\end{equation}

Therefore, the classification function of VQC is realized through the geometric relationship of the two classification lines $y+{k_1}x=0$ and $y+{k_2}x=0$, where $k_1 = \frac{\sin(2\theta + \frac{\pi}{4})}{\sin(2\theta - \frac{\pi}{4})}$ and $k_2 = -\frac{\sin(2\theta - \frac{\pi}{4})}{\sin(2\theta + \frac{\pi}{4})}$. When $(y+k_1x)\cdot(y+k_2x)\ge0$ we assign the predicted label to 0, otherwise 1. It is obvious that $k_1 \cdot k_2 = -1$, indicating the two classification lines are mutual perpendicular. The expressibility of VQC is limited by this vertical characteristic due to the independence of the model parameter. 

Now we will prove that, due to the property of Pattern 1, VQC can theoretically achieve a $100\%$ maximum classification accuracy with sufficient training data set and training epoch. However, for Pattern 2 and Pattern 3 whose geometric property is not a type of vertical margin, VQC can only achieve a maxmimum classification accuracy much lower than $100\%$. 

As shown in Fig.~\ref{fig-ability},  the orange solid lines with slope angle ${\beta}_1$ and ${\beta}_2$ represent the actual dividing margin of a certain pattern. And the green dashed lines with slope angle ${\gamma}_1$ and ${\gamma}_2$ refer to the classification lines at current model parameters. The blue arrows indicate the correctly classified zone and the red arrows point to the error zone. Two situations may arise during the training process. We first consider a general situation given in Fig.~\ref{fig-ability}~(a), where two classification lines cross through data points in the same class. In this situation, data items of at least one class are exactly classified, and the maxmimum classification accuracy is
\begin{equation}\label{eqn:max-accuracy}
\begin{aligned}
acc &= 1 - \underset{\gamma_1, \gamma_2} {\text{min}} \frac{||\beta_1 - \beta_2| - |\gamma_1-\gamma_2||}{\pi}
& = 1 - \underset{\theta} {\text{min}} \frac{||\beta_1 - \beta_2| - \arctan|\frac{k_1-k_2}{1+k_1k_2}||}{\pi}
\end{aligned}.
\end{equation}

For the other situation shown in Fig.~\ref{fig-ability} (b), where two classification lines cross through data points in different classes, the classifier can easily move the situation to the previous case through adjusting the value of parameter single-qubit gate (HWP($\theta$)), so Equation~(\ref{eqn:max-accuracy}) still holds.

As for VQC, there is a constrain between ${\gamma}_1$ and ${\gamma}_2$ due to the perpendicular relationship of two classification lines, which can be expressed as ${\gamma}_1 = {\gamma}_2 + \frac{\pi}{2}$. Therefore, the classification accuracy $acc = 1 - \frac{||\beta_1 - \beta_2| - \frac{pi}{2}|}{\pi}$, implying that the maxmimum classification accuracy totally depends on the proportion of each class for a certain pattern. As is shown in Fig.~2(a) in the maintext, the slope angle difference $|\beta_1 - \beta_2|$ of the actual dividing margins for Pattern 1-3 are $\pi/2$, $\pi/4$ and $\arctan(\frac{1}{4})$ respectively. Then we can easily deduce a maxmimum accuracy of 100\% for Pattern 1, 75\% for Pattern 2 and 57.79\% for Pattern 3, which is validated in the experiments in the maintext.

\subsection{Nonlinearity-Enhanced Variational Quantum Classifier}
\begin{figure*}
\begin{center}
\includegraphics[width=0.8\linewidth]{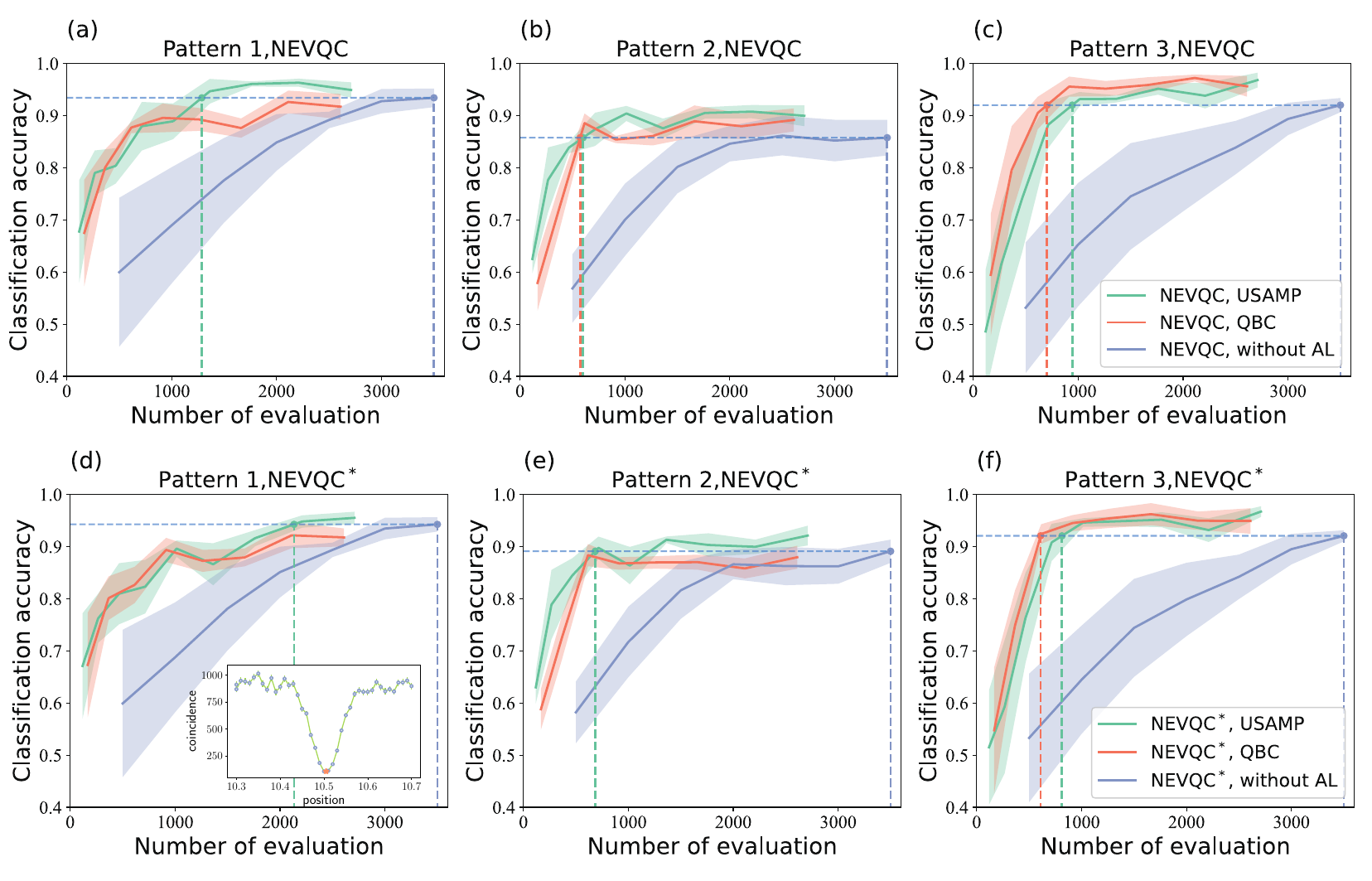}
\end{center}
\caption{\textbf{The performance of NEVQC and NEVQC$^*$ with and without AL strategies during the whole training process.} Each line shows an average result over four independent experiments and the bands show the standard deviation. The blue dashed lines mark the classification accuracy of NEVQC and NEVQC$^*$ (without AL) at convergence and the corresponding numbers of evaluations. The green and orange dashed lines show the the required numbers of evaluations of NEVQC and NEVQC$^*$ (with USAMP and QBC, respectively) that achieve the same classification accuracy as that without AL. In Subfig.~(d), we also show the process of searching the interference point by moving the position of prism (The corresponding device setup is shown in Fig.~2c in the maintext.). }
\label{fig-coherence} 
\end{figure*}

The nonlinearity-enhanced variational quantum classifier~(NEVQC) is realized by introducing an ancilla qubit (prepared as $\ket{+}$ state), a PBS operation, and post-selection on the ancilla qubit after a parameterized single-qubit gate (HWP($\theta_2$)) applied on it. We note that the  implementation of NEVQC does not require the interference of two photons on the PBS, i.e. two photons do not require space-time encounters on the PBS. In this case, after postselecting the events where there is exactly one photon exiting each output of the PBS, the quantum state is 

\begin{equation}\label{eqn:PBS}
(\alpha\ket{0}+\beta\ket{1})\ket{+}\xrightarrow[\text{no interference}]{\text{PBS}+\text{postselection}}\frac{1}{\sqrt{2}}(|\alpha|^2\ket{00}\bra{00}+|\beta|^2\ket{11}\bra{11}).
\end{equation}
where 

$\alpha\ket{0}+\beta\ket{1}$ is the data qubit after applying the first parametered single-qubit gate (HWP$(\theta_1)$). We also implement the same experiments for the NEVQC with interference occurance on the PBS, which we name as NEVQC$^*$. As shown in Fig.~2(c) in the maintext, we put a triangular prism on the path of the data qubit. The electric linear translation stage below it can be used to adjust the optical path difference between the two photons. We scan the position of the triangular prism to check the indistinguishability of the outcome photon. When the triangular prism reaches the interference point, the opration becomes
\begin{equation}
(\alpha\ket{0}+\beta\ket{1})\ket{+}\xrightarrow[\text{interference}]{\text{PBS}+\text{postselection}}\frac{1}{\sqrt{2}}(\alpha\ket{00}+\beta\ket{11}).
\end{equation}
which is different from the condition that no interference is happened on the PBS.

Now we will show that, although the transformation formula is distinct, we still end up with the same probability expression of $|0\rangle$(or $|1\rangle$)events in the input qubit. We take density matrix representation for NEVQC to better describe the transformation.
\begin{equation}
\begin{aligned}
\rho^{\text{NEVQC}}_{\text{en}} &= |\psi_0\rangle^d |0\rangle^c \langle \psi_0|^d \langle 0|^c \\
&\xrightarrow[{\text{H}^c}]{\text{HWP}(\theta_1)^d} |\psi_1\rangle^d|+\rangle^c \langle \psi_1|^d \langle +|^c \\
&\xrightarrow[]{\text{PBS}} ({x_i}\cos2{\theta}_1 + {y_i}\sin2{\theta}_1)^2 |00\rangle\langle 00| + ({x_i}\sin2{\theta}_1 - {y_i}\cos2{\theta}_1)^2 |11\rangle\langle 11| \\
&\xrightarrow[\text{post-select} |0\rangle^c \text{on ancilla}]{\text{HWP}(\theta_2)^c} (\cos2{\theta}_2({x_i}\cos2{\theta}_1 + {y_i}\sin2{\theta}_1))^2 |0\rangle^d \langle 0|^d + (\sin2{\theta}_2({x_i}\sin2{\theta}_1 - {y_i}\cos2{\theta}_1))^2 |1\rangle^d \langle 1|^d,
\end{aligned}
\end{equation}
where $ |\psi_0\rangle^d = (x_i|0\rangle^d + y_i|1\rangle^d $ is the data qubit, $|0\rangle^c$ is the ancilla qubit, and $|\psi_1\rangle^d = ({x_i}\cos2{\theta}_1 + {y_i}\sin2{\theta}_1) |0\rangle^d+ ({x_i}\sin2{\theta}_1 - {y_i}\cos2{\theta}_1) |1\rangle^d $.

Similarly, we can deduce the transformation in NEVQC$^*$ as
\begin{equation}
\begin{aligned}
|\psi\rangle^{\text{NEVQC}^*}_{\text{en}} &= |\psi_0\rangle^d |0\rangle^c\\
&\xrightarrow[{\text{H}^c}]{\text{HWP}(\theta_1)^d} |\psi_1\rangle^d|+\rangle^c \\
&\xrightarrow[]{\text{PBS}} ({x_i}\cos2{\theta}_1 + {y_i}\sin2{\theta}_1) |00\rangle + ({x_i}\sin2{\theta}_1 - {y_i}\cos2{\theta}_1) |11\rangle \\
&\xrightarrow[\text{post-select} |0\rangle^c \text{on ancilla}]{\text{HWP}(\theta_2)^c} (\cos2{\theta}_2({x_i}\cos2{\theta}_1 + {y_i}\sin2{\theta}_1)) |0\rangle^d  + (\sin2{\theta}_2({x_i}\sin2{\theta}_1 - {y_i}\cos2{\theta}_1)) |1\rangle^d.
\end{aligned}
\end{equation}

Therefore, the output probability of $\ket 0$ and $\ket 1$ is exactly the same for the NEVQC and NEVQC$^*$, although the final states of NEVQC and  NEVQC$^*$ are mixed state and pure state, respectively. Fig.~\ref{fig-coherence} shows the comparison of experimental results between NEVQC and NEVQC$^*$. We observe that the two classifiers have almost identical performance for classification on three patterns, which is consistent to our theoretical derivation.

In the next part, we will present the maximum classification accuracy for NEVQC and NEVQC$^*$. Similar to the analysis of VQC, assuming $\braket{Z}>0$, we have
\begin{equation}
\begin{aligned}
(\cos2{\theta}_2({x_i}\cos2{\theta}_1 + {y_i}\sin2{\theta}_1))^2 > 
(\sin2{\theta}_2({x_i}\sin2{\theta}_1 - {y_i}\cos2{\theta}_1))^2.
\end{aligned}
\end{equation}

Then
\begin{equation}
\begin{aligned}
({x_i}\cos(2{\theta}_1-2{\theta}_2)+{y_i}\sin(2{\theta}_1 - 2{\theta}_2))
\cdot
({x_i}\cos(2{\theta}_1+2{\theta}_2)+{y_i}\sin(2{\theta}_1 + 2{\theta}_2))>0
\end{aligned}.
\end{equation}

We notice that two classification lines of NEVQC have slopes $k_1 = \cot(2{\theta}_1 - 2{\theta}_2)$ and $k_2 = \cot(2{\theta}_1 + 2{\theta}_2)$, with an included angle 
\begin{equation}
\begin{aligned}
|\gamma_1 - \gamma_2| = \arctan(|\frac{k_1 - k_2}{1+k_1{k_2}}|) = \arctan(|\tan(4{\theta}_2)|)
\end{aligned}.
\end{equation}

Using Equation~(\ref{eqn:max-accuracy}), we get the maximum classification accuracy 
\begin{equation}\label{eqn:max-accuracy2}
\begin{aligned}
acc &= 1 - \underset{\theta_1, \theta_2} {\text{min}} \frac{||\beta_1 - \beta_2| - \arctan|\frac{k_1-k_2}{1+k_1k_2}||}{\pi}
&= 1 - \underset{\theta_2} {\text{min}} \frac{||\beta_1 - \beta_2| - \arctan|\tan(4\theta_2)|}{\pi}.
\end{aligned}
\end{equation}

Although $\theta_1$ doesn't contribute to the maxmimum classification accuracy, it ensures that two classification lines can always transfer to the fisrt classification situation shown in Fig.~\ref{fig-ability}. Along with the training process, NEVQC iteratively adjusts the slope angles of classification lines while optimizing their included angle. Hence, the model expressibility of NEVQC is well enhanced. According to Equation~(\ref{eqn:max-accuracy2}), there exists at least one solution $\theta_2 = \frac{|\beta_1 - \beta_2|}{4}$ to achieve a maximum classification accuracy 100\% for any pattern. That implies NEVQC theoretically outperforms its single-qubit counterpart VQC.

\section{classical simulation}
\subsection{Numerical simulation}
We perform abundant numerical simulations to support our experimental settings. In our active learning experiments, the classifier iteratively selects 10 data items from an unlabeled data pool with size 20. After a newly chosen data item is added, the classifier trains itself for 10 epochs with its labeled training dataset. As for training process without active learning, we just train the classifier for 35 training epochs with 20 labeled training dataset. 

\begin{figure*}[t]
\begin{center}
\includegraphics[width=0.8\linewidth]{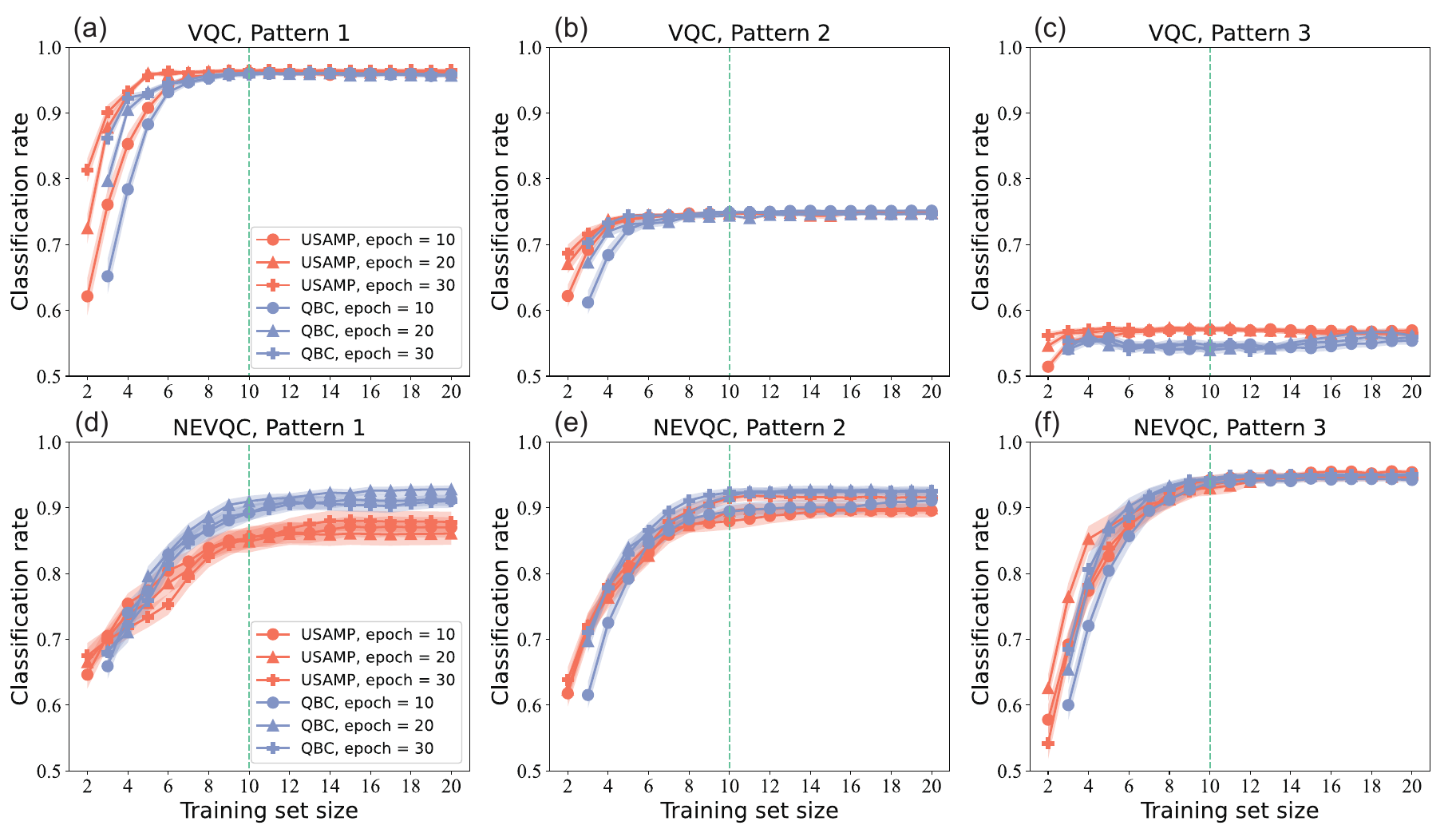}
\end{center}
\caption{\textbf{comparison among different training epoch after a newly chosen data item added. }The classification accuracy plots show the performance of classifier with training epoch 10, 20 and 30. We iteratively increase the labeled training set size up to 20. The vertical green dashed line refers to the position where 10 data items have been added to the labeled training data pool.}
\label{fig-epoch}
\end{figure*}

The rationality and efficiency of the experimental settings are demonstrated by numerical simulations. 
For each classifier and data pattern, 100 independent numerical experiments are implemented to calculate the mean and standard deviation of the results. Fig.~\ref{fig-epoch} shows the comparison among different training epoch after a newly chosen data item added. During the training process, we iteratively increase the labeled training set size up to 20. We can see that for training epoch 10, 20 and 30, the classification accuracy curve reaches the final convergence with little difference. The green dashed line indicates the position where 10 data items have been labeled to train the classifier. It is obvious that 10 labeled data items are sufficient for training the classifier close to its maxmimum performance. Therefore, considering the balance between classification performance and training time consumption, we finally set the number of selected data items to 10. And for each newly added item, we train the classifier for 10 training epoch with its current labeled dataset.
\subsection{Experimental performance compared to numerical simulation}

\begin{figure*}[t]
\begin{center}
\includegraphics[width=0.7\linewidth]{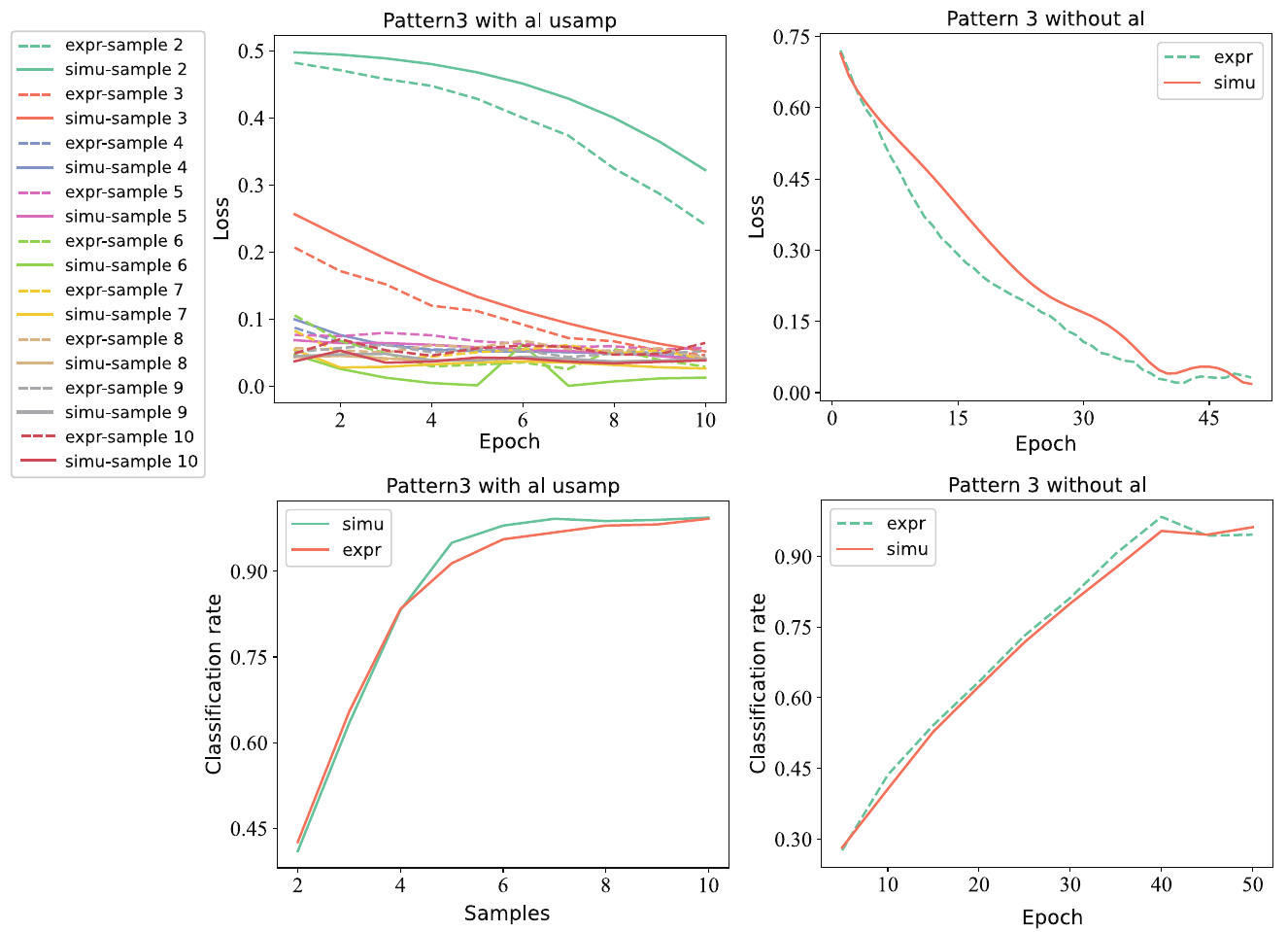}
\end{center}
\caption{\textbf{Detailed comparison for experimental and simulated results of one set of experiments. }The pictures in the upper column show the decline curves of loss, and those in the lower demonstrate the classification accuracy on the test dataset. We choose one set of experimental and simulated results from NEVQC on Pattern 3, with and without active learning, respectively. }
\label{fig-comparison}   
\end{figure*}

\begin{figure*}[htbp]
\begin{center}
\includegraphics[width=0.8\linewidth]{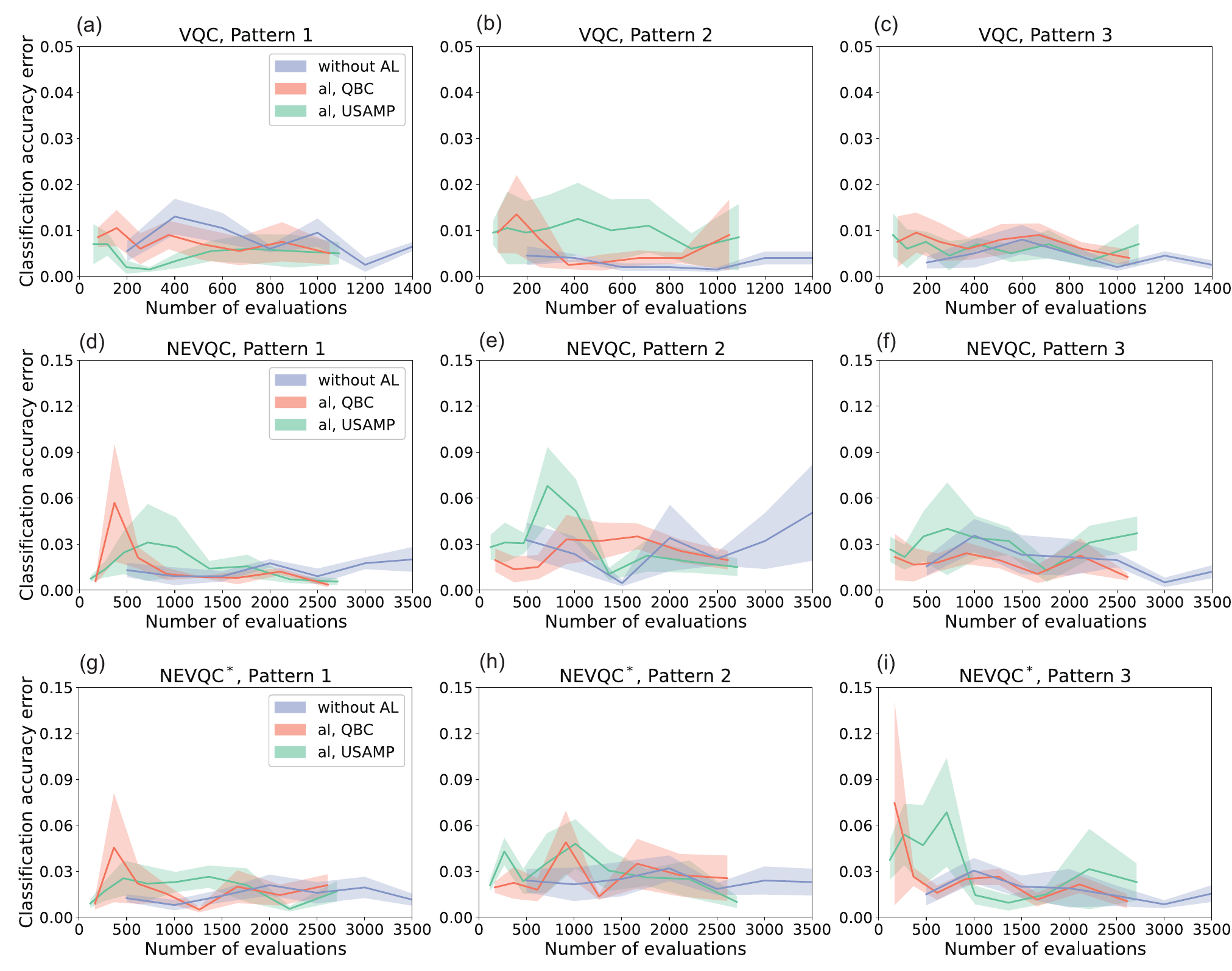}
\end{center}
\caption{\textbf{The average absolute classification accuracy error bewteen experiments and their simulations. } For active learning process, the absolute classification accuracy error is calculated every time a newly chosen data item is added to the labled data pool. As for training without active learning, we test the model performance on test dataset every five training epoch.}
\label{fig-accuracy-error}
\end{figure*}

\begin{figure*}[t]
\begin{center}
\includegraphics[width=0.8\linewidth]{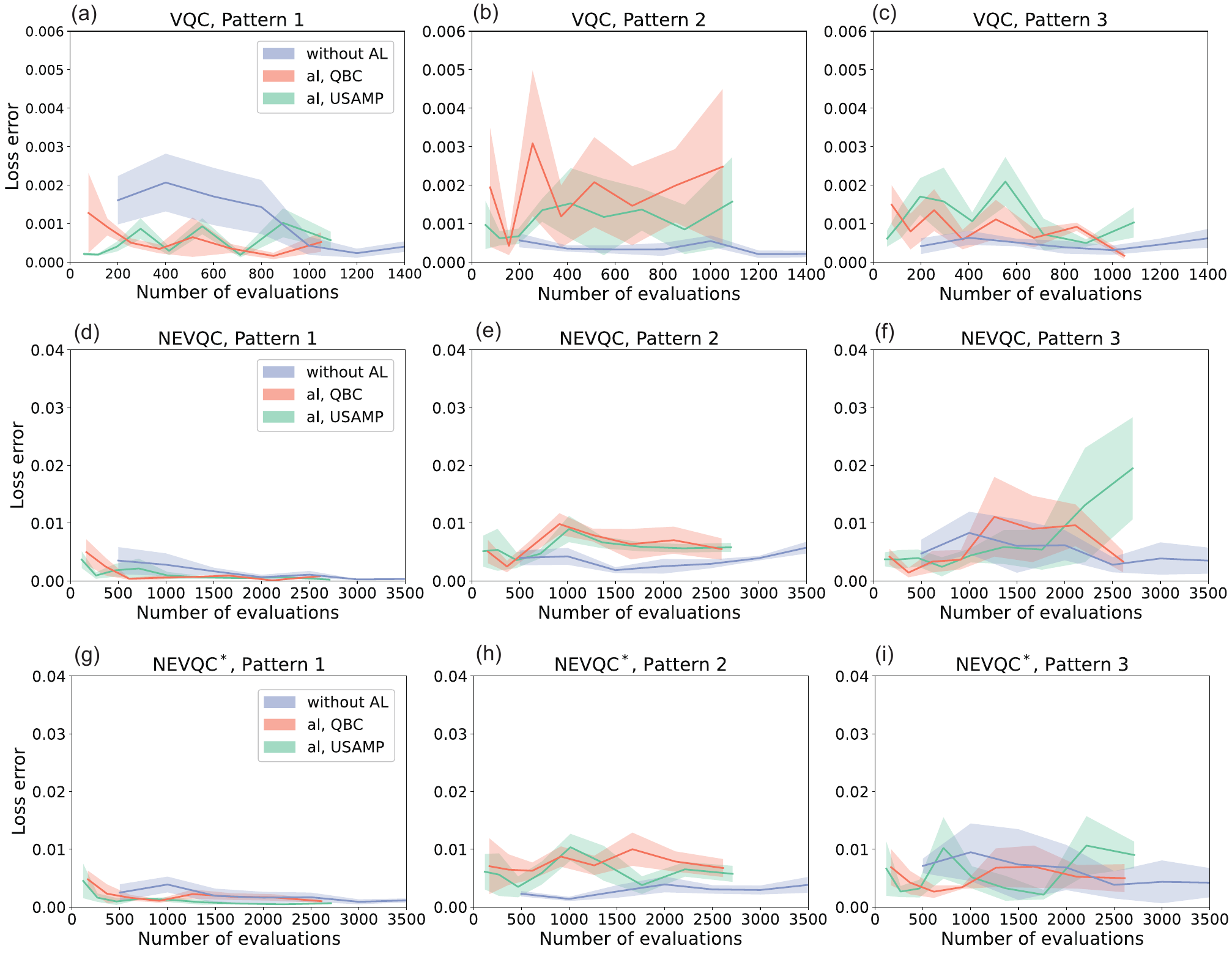}
\end{center}
\caption{\textbf{The average absolute loss error bewteen four sets of experiments and their simulations. }The way to collect and display the comparison results is similar to that of classification accuracy error.}
\label{fig-loss-error}
\end{figure*}

We also compare our experimental outcomes to numerical simulations, to show the high-precision manipulation of our experiments. For experiments and simulations, we take the same initial data set and parameter values. In Fig.~\ref{fig-comparison}, we show the comparison result of one set of experiments from NEVQC on Pattern 3, with active learning and without active learning respectively. We can see that the experimental results fit their simulated data well. The difference between theoretical and experimental results appears to be small, which contains both experimental noise and statistical error.

Furthermore, in order to quantify the average performance of the whole experiments, we present the average absolute loss and accuracy error of experiments by comparing to their theoretical simulations in Fig.~\ref{fig-accuracy-error} and Fig.~\ref{fig-loss-error}. We compute the average absolute loss and accuracy error using the following equation.
\begin{equation}
\begin{aligned}
\text{Error} = \frac{1}{L}\sum_{\text{i}=1}^L |v_{i}^{expr} - v_{i}^{simu}|,
\end{aligned}
\end{equation}
where $v_{i}^{expr}$ and $v_{i}^{simu}$ represent the value of loss or classification accuracy of experiments and simulations, and $L$ refers to the training data size (or test data size) for loss error (or classification accuracy error). For active learning process, the absolute classification accuracy error is calculated each time a newly chosen data item is added to the labled data pool. As for training without active learning, we test the model performance on test dataset every five training epoch. Four independent sets of experiments are implemented for the three classifiers (VQC, NEVQC and NEVQC$^*$ mentioned in the maintext) to report the average value and the standard deviation. 

As can be seen, there is a good agreement between the experimental results and simulations, both for VQC and NEVQC. Furthermore, we notice that VQC has shown better experimental accuracy and stability, with an upper bound 0.01 for classification accuracy error and 0.005 for loss error. This phenomenon is understandable because VQC uses fewer qubits and quantum operations than NEVQC. 


\section{Demonstration of the Training process with AL strategy}
\begin{figure*}
\begin{center}
\includegraphics[width=0.8\linewidth]{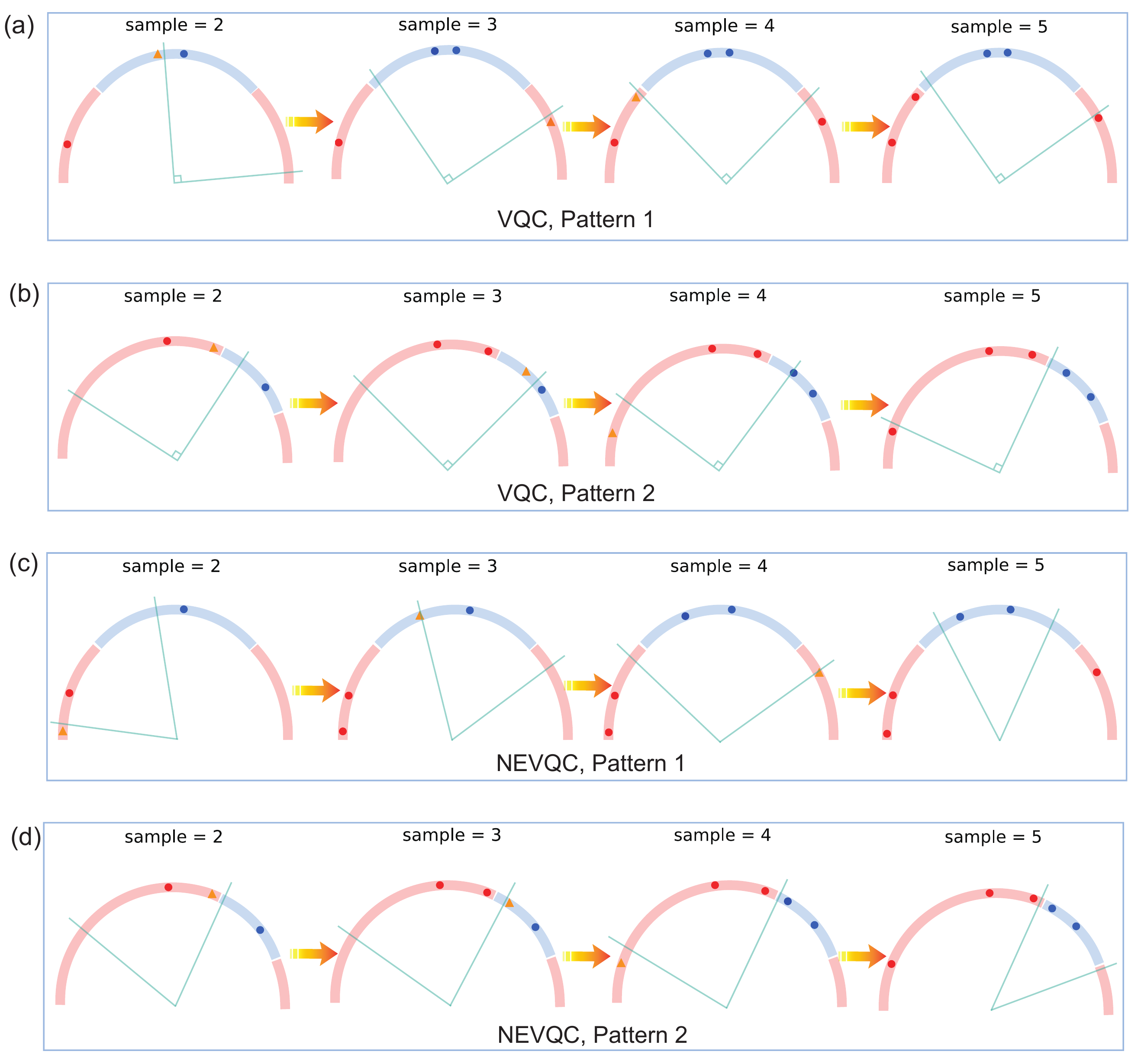}
\end{center}
\caption{\textbf{The active learning training process of (a-b) VQC and (c-d) NEVQC using USAMP strategy.} We choose one set of experimental results to demonstrate the training process. The classifier iteratively adds data item to the labeled data pool and trains itself. We record the model parameters and draw the current training data set and classification lines in the pictures. The red and blue circles respectively refer to the data items of class 1 and class 0 in the labeled data pool, and the newly chosen data item in the next round is represented by an orange triangle.}
\label{fig-training}
\end{figure*}

The active learning training process of VQC and NEVQC (both using USAMP strategy) on Pattern 1 and Pattern 2 is given in Fig.~\ref{fig-training}. We choose one set of experimental results to demonstrate the actual training process. The classifier iteratively adds data item to the labeled data pool and trains itself. We record the model parameters and draw the training data set and classification lines in Fig.~\ref{fig-training}. We set the initial training data set by randomly choose two data items with different labels. The blue and red dots respectively represent the data points in class 1 and class 0, and the orange triangle denotes the newly chosen data item in the next epoch.

We can see that as the training data size increases, the model classification lines are trained to get closer to the actual dividing margins, providing a more intuitive understanding of the model ability mentioned in previous section. Fig.~\ref{fig-training} (a-b) show that the classification lines of VQC are always maintaining a vertical relationship while optimization. But for the NEVQC given in Fig.~\ref{fig-training} (c-d), the included angle of two classification lines is continuously optimized as training data size increases. The classification lines are optimized to cross through data points in the same class, which supports the derivation of Equation~(\ref{eqn:max-accuracy}).

Moreover, during the active learning training process, the newly chosen data item is always close to the model classification line at current model parameters, which is on great agreement with the least confidence principle in USAMP.

\section{Probability vanishment in post-selection}
When there are no $|0\rangle$ events happened in the ancilla in NEVQC, we call this situation probability vanishment. Probability vanishment is an instinctive phenomenon rather than an erroneous occurance. As for NEVQC, the probability of $|0\rangle$ events in the ancilla is
\begin{equation}
\begin{aligned}
\text{P}_0^* = (\cos2{\theta}_2({x_i}\cos2{\theta}_1 + {y_i}\sin2{\theta}_1))^2 + (\sin2{\theta}_2({x_i}\sin2{\theta}_1 - {y_i}\cos2{\theta}_1))^2,
\end{aligned}
\end{equation}
there always exists some singularity such that the probability value $\text{P}_0^*$ turns to zero, e.g. when ${\theta}_2 = 0$ as well as direction vectors $(x_i, y_i)$ and $(\cos2{\theta}_1, \sin2{\theta}_1)$ are orthogonal. 
In the experiments, since the $Z$ expectation is measured using finite number of coincidence events among Detector 4-6
\begin{equation}
\braket{Z}=\frac{N_{45}-N_{46}}{{N_{45}}+N_{46}-2N_{456}},
\end{equation}
When the parameter point nears to those singularities, the probability $\text{P}_0^*$ becomes so negligible that we can not observe any $|0\rangle$ event in the experiment. The denominator of the calculation formula then turns to zero. This exception could indeed interrupt our experiments, because we can not avoid the vanishment from happening, or extract information of the expectation through other approaches. 

Therefore, once probability vanishment occurs, we skip the post-selection and set $\braket{\text{Z}(x_i,\theta)}$ to zero, the unbiased estimation of the expectation assuming its uniform distribution in the interval $[-1, 1]$. During our experiments, we observe that probability vanishment is a rather occasional phenomenon and the experimental results are in good agreement with our numerical simulations, as is shown in Fig.~\ref{fig-accuracy-error} and Fig.~\ref{fig-loss-error}.

\bibliographystyle{apsrev4-2}

\bibliography{b}